\documentclass[review,onefignum,onetabnum]{siamart190516}

\usepackage{xr}

 \usepackage{amsfonts}
\usepackage{graphicx}
\usepackage{epstopdf}
\usepackage{algorithmic}
\ifpdf
  \DeclareGraphicsExtensions{.eps,.pdf,.png,.jpg}
\else
  \DeclareGraphicsExtensions{.eps}
\fi

\newsiamremark{remark}{Remark}
\newsiamremark{hypothesis}{Hypothesis}
\crefname{hypothesis}{Hypothesis}{Hypotheses}
\newsiamthm{claim}{Claim}

\headers{StringNET}{J. Han, S. Gu and X. Zhou}

\title{StringNET: Neural Network based Variational Method for Transition Pathways}

\author{Jiayue Han\thanks{School of Data Science, City University of Hong Kong, Tat Chee Ave, Kowloon, Hong Kong SAR.}
\and Shuting Gu\thanks{College of Big Data and Internet, Shenzhen Technology University, Shenzhen 518118, P.R. China}
\and Xiang Zhou \thanks{Corresponding author. School of Data Science and Department of Mathematics, City University of Hong Kong, Tat Chee Ave, Kowloon, Hong Kong SAR 
(xizhou@cityu.edu.hk).}}

\usepackage{amsopn}

\usepackage{graphicx}
\usepackage{dcolumn}
\usepackage{bm}
\usepackage[mathlines]{lineno}
\linenumbers\relax 

\usepackage[utf8]{inputenc}
\usepackage[T1]{fontenc}
\usepackage{mathptmx}
\usepackage{etoolbox,comment}
\usepackage{subfigure}
\usepackage{color}

\usepackage{array}

\newtheorem{thm}{Theorem}

\usepackage{todonotes}

\graphicspath{{figures/}}

\usepackage{physics}

\usepackage{hyperref}
\hypersetup{
     unicode=false,          
    pdftoolbar=true,        
    pdfmenubar=true,        
    pdffitwindow=false,     
    pdfstartview={FitH},    
    pdftitle={My title},    
    pdfauthor={Author},     
    pdfsubject={Subject},   
    pdfcreator={Creator},   
    pdfproducer={Producer}, 
    pdfkeywords={keyword1} {key2} {key3}, 
    pdfnewwindow=true,      
    colorlinks=true,       
    linkcolor=red,          
    citecolor=green,        
    filecolor=magenta,      
    urlcolor=cyan           
}

\newcommand{\inpd}[2]{\left\langle #1, #2 \right\rangle}

\newcommand{\wt}[1]{\widetilde{#1}}

\newcommand{\Real}{\mathbb {R}}

\newcommand{\eps}{\varepsilon}

\renewcommand{\d}{\mathop{}\!\ensuremath{\mathrm{d}}}

\newcommand{\dt}{ ~\ensuremath{\mathrm{d} t } }
\newcommand{\dx}{ ~\ensuremath{\mathrm{d} x} }

\newcommand{\xb}{\mathbf x}

\newcommand{\e}{ \mathbb{E}}

 \newcommand{\pll}{\kern 0.56em/\kern -0.8em /\kern 0.56em} 

\begin{document}
\nolinenumbers
\maketitle

 \begin{abstract}
Rare transition events in meta-stable systems under noisy fluctuations are crucial for many non-equilibrium physical and chemical processes. In these processes, the primary contributions to reactive flux are predominantly near the transition pathways that connect two meta-stable states. Efficient computation of these paths is essential in computational chemistry. In this work, we examine the temperature-dependent maximum flux path, the minimum energy path, and the minimum action path at zero temperature. We propose the StringNET method for training these paths using variational formulations and deep learning techniques. Unlike traditional chain-of-state methods, StringNET directly parametrizes the paths through neural network functions, utilizing the arc-length parameter as the main input. The tasks of gradient descent and re-parametrization in the string method are unified into a single framework using loss functions to train deep neural networks. More importantly, the loss function for the maximum flux path is interpreted as a softmax approximation to the numerically challenging minimax problem of the minimum energy path. To compute the minimum energy path efficiently and robustly, we developed a pre-training strategy that includes the maximum flux path loss in the early training stage, significantly accelerating the computation of minimum energy and action paths. We demonstrate the superior performance of this method through various analytical and chemical examples, as well as the two- and four-dimensional Ginzburg-Landau functional energy.
 \end{abstract}

\begin{keywords}
  rare event, transition paths; maximum flux path; 
  minimum energy path; neural networks;  minimum action method;
\end{keywords}

\begin{AMS}
65K05, 65K10, 68T07, 82B26, 49S05,


\end{AMS}

\section{Introduction }
\label{sec:level1}

The study of the transition phenomenon between meta-stable states on the energy landscape of molecular systems    is crucial to understanding
the persistent impact of random fluctuations over a
long time scale   in physical, chemical and biological sciences. The transition process is generally described by various notions of  paths between  local minima of  the potential energy function  \cite{E2005242,FECbook2007}.
Particularly,  in the small noise limit,  the minimum energy path (MEP) has been extensively studied 
on the energy surface in gradient systems. Based on the large deviation theory \cite{FW2012}, the minimum action path (MAP) is of greatest interest for a general dynamical system perturbed by small noise.  
 
Numerous computational techniques have been devised to determine the minimum energy paths, including the nudged elastic band (NEB) method \cite{NEB1998,NEBtanget2000} and the (zero temperature) string method \cite{String2002, String2007}, both of which    optimize a finite number of discrete points (``images'') used to represent  the path. Thus,  these methods are generally referred to as the chain-of-state  methods. In the NEB, the total energy 
of these images is minimized with virtual spring  forces to enforce the proper  curve parametrization.  The string method likewise evolves the images on the string first independently then 
performs the important curve reparametrization with the aid of numerical interpolation. For non-gradient systems, the minimum action method (MAM) \cite{weinan-MAM2004,KS-WZE2009} for  a fixed time interval is  based on the large deviation principle  \cite{FW2012}, and has been developed with a crucial improvement of the geometric method based on  the Maupertuis  principle -- geometric Minimum Action Method (gMAM) \cite{Heymann2006,Heyman2008}. 
These path-finding algorithms are already in the form of a variational problem by minimizing the action functional in the path space. For the energy functional in Hilbert space   such as the phase field models, the numerical computation usually solves the  corresponding Euler-Lagrangian equation in the form of boundary-value problem.

For the finite temperature case where the noise amplitude cannot be ignored, the transition path sampling \cite{TPSChandler2002} is a classic method
but it has a very large computational burden.
The transition path theory (TPT)  \cite{Weian-TPT-2006,TPT_review,TPTICIAM} offers a theoretical foundation and can recover many existing methods under various assumptions. 
The maximum flux path has been proposed earlier   \cite{Berkowitz1983} and can be rigorously derived under the transition path theory by the  localized assumption 
in form of a  path.   The maximum flux path is the pathway that maximizes  the probability flux  between two metastable states, hence capturing the most probable route that the system will take during its transition.  The underlying variational characterization of the maximum flux path caters for numerical optimization in the path space and  thus circumvents  the much   more expensive  computation of the  committor function in the original TPT. The numerical computation of this path offers a powerful tool for identifying the most likely transition pathways in noisy environments. 
Furthermore, the maximum flux path can be contrasted with other paths such as the minimum energy path (MEP) and the minimum action path (MAP). While the MEP is concerned with the energetically favorable route and the MAP with the path that minimizes the action functional, the maximum flux path focuses on the probabilistic aspect, making it particularly relevant for systems at finite temperatures where thermal fluctuations cannot be ignored.
 
In the computational aspects,
the numerical computation of these paths
is dominantly based on the chain-of-state
method by discretizing the path into a sequence of  images and evolving each individual image.
Recently,   the emergence of using other numerical techniques  neural networks(NNs)  has revolutionized
  many scientific areas, particularly for 
high dimensional equations \cite{e_deep_2018,han2018solving}.  This type of neural network representation and the 
training flexibility of loss functions have also spurred the 
application to the path calculation.
The geometric minimum action path has recently been  investigated   by a deep learning approach  \cite{SIMONNET2023112349}
for   general non-gradient systems.  We focus on the gradient system here, although we also use the geometric action as the loss function. The corresponding Euler–Lagrange equations for the minimizing 
   paths of the Onsager–Machlup   and the Freidlin–Wentzell  functionals
 have been solved    by   neural networks  with a least squared loss  \cite{CHEN2023133559} in the fashion of the widely
 used Physics Informed Neural Network.  Our work here exploits the  variational formulation for the pathways instead of focusing on the first order necessary condition for optimality. 
  Additionally, the most probable transition path problem is regarded as an optimal control problem and also computed  by the neural network method, particularly for   non-Gaussian noise \cite{DuanPathOCChaos2022}.

In this paper, we propose the StringNET method which adopts the neural networks for the path calculation to compute the maximum flux path and the minimum energy path in gradient systems.
In contrast to the chain-of-state methods such as the string method,
the StringNET is based on the the corresponding variational formulations and the  efficient training methods for the neural networks representing the path.
We shall construct  three loss functions to characterize
the maximum flux path (at any given temperature), the minimum energy path and the minimum action path (for gradient system). 
The arc length parametrization of the path  is naturally enforced by the additional loss function as a penalty term.


 The features and contributions of this paper include  the following. 
 \begin{enumerate}
     \item  By using the non-parametric path representation, the discretization  and   interpolation error  between images in traditional chain-of-states methods are eliminated in principle and the output is a continuous $\Real^d$-valued  function of the arc length parameter $s$.
     The deep neural network can directly handle the path in the Hilbert space such as $L^2(\Real^n)$.
We also validate   that   using one neural network for the path is not only easier to code but also more   efficient    in training 
than using totally $d$  networks for each component of the path.

     \item 
    We rigorously show   that the loss for   
maximum flux path converges to the loss for minimum energy path, as the temperature vanishes.   We empirically find that the training of neural networks for the max-flux path is the easiest  since the loss function does not involve the potential gradient. 
We then propose the pre-training method for computing the MEP  by incorporating the loss of max-flux path with decaying weight.  This not only accelerates the training speed  but also enhances the numerical 
robustness in challenging cases.

\item   We conduct numerical experiments  for paths in both Euclidean space  and  the Hilbert space. The examples include the classic chemistry example and the 
high dimensional Ginzburg-Landau functional.
We  carefully compare the performance of the loss functions 
and the network structures on these examples. We reach the important
observation for real applications that  
the StringNET with the loss $\ell_\beta$ based on the maximum flux, and the loss $\ell_g$ based on the geometric action minimization, works most efficiently.
 \end{enumerate}

The rest of the paper is organized as follows: In Section \ref{Problem_formulation}, we will introduce the formulation of the problems and the definitions of paths.
Then the corresponding numerical details of StringNET method  are discussed in Section \ref{main_method}. Section \ref{sec:NT} demonstrates the performance of the StringNET method in various numerical examples. The   paper is concluded in Section \ref{sec:C}.

\section{\label{Problem_formulation} Models and Path formulations}

\subsection{ Models and Notations }

We are concerned with the paths arising from the most probable transition events  in the small-noise perturbed 
gradient dynamical system
in the form of over-damped Langevin equation: 
\begin{equation} \label{SDE}
{\rm d}X(t) = -\nabla U(X(t)) {\rm d}t + \sqrt{2\beta^{-1}} {\rm d}W(t).
\end{equation}
where $X(\cdot):
\mathbb{R}_+ \to \mathbb{R}^d$ is the particle position   at time $t$,  $U(\cdot)\in  C^2(\mathbb{R}^d; \Real)$ is the potential energy function and $W(\cdot)$ is the Brownian motion. $\beta=1/k_B T$   is the inverse temperature,
$k_B$ is the Boltzmann constant, 
and a large value of $\beta$ corresponds to a lower temperature.  

Consider two local minimum points of the function $U$, denoted by 
$a$ (the reactant configruation) and $b$ (the product configuration).
Let $AC_{a,b}([0,1]; \Real^d)$ be the space of  absolute continuous function $\varphi(s)$ satisfying $\varphi(0)=a$ and $\varphi(1)=b$.
This  AC space is equipped with the norm $\|f\|_{AC}=\|f\|_\infty + \int_0^1 |f'(t)| \d t$.
We will be  interested in the   three types of paths   in this path space $AC_{a,b}$ to be specified later. All paths are defined in geometrically, meaning free of parametrization. So without loss of generality, we  adopt  the arc-length parametrization   which  means that 
$\int_0^t   
\abs{ \varphi'(s)} \d s = t \int_0^1  
\abs{ \varphi'(s)} \d s $ for all $t\in [0,1].$
Here $\varphi'(s)=\partial_s \varphi(s)$ is the derivative 
and $\abs{\bullet}$ is the norm  on the tangent space $\Real^n$, i.e., $\abs{\bullet}$ is the usual Euclidean norm in $\Real^n$.
$\inpd{\cdot}{\cdot}$ is then refers to the inner product.
If one extends the above formulation to the path space on   Riemannian manifold
(e.g.,  Hilbert space),  $\inpd{\cdot}{\cdot}$ (resp. $\abs{\bullet}$) becomes the corresponding Riemannian metric tensor (resp. metric).
$\int_L \bullet ~\d\varphi =\int_0^1 \bullet~\abs{\varphi'(s)}\d s$ indicates the line integral, which is invariant with reparametrization of the curve $\varphi$.

\subsection{  Paths and Variational Forms}

{\bf Minimum Energy Path.}~
The Minimum Energy Path (MEP) is a concept used   in the study of reaction dynamics and transition state theory. It represents   the pathway that requires the least amount of energy barrier, for a system to transform from reactants to products during a chemical reaction. 

We first regard the MEP  with  the minimax principle    in  the Mountain Pass Theorem \cite{BisgardSIAMREV2015}, which  provides a mathematical justification for the existence of saddle point under certain conditions (such as Palais–Smale compactness condition).
Consider the following  variational  problem arising from the mountain pass theorem:
\begin{equation}\label{minmaxF}
 \min_{\varphi\in AC_{a,b}}  \norm{U\circ \varphi}_\infty =  \min_{\varphi\in AC_{a,b}} ~ \max_{0\le s \le 1} U(\varphi(s)).
\end{equation}
where $AC_{a,b}$ is a short notation for  $AC_{a,b}([0,1];\Real^d)$.
This is a minimax problem and is  very challenging to solve directly  due to   the maximum norm.
In addition, the minimizing path   in Equation \eqref{minmaxF}  does not exhibit uniqueness. Any path that traverses the lowest saddle point between   
$a$ and $b$, and subsequently descends in potential value when away from the saddle point, can be considered a minimizing path. This descent path does not need to strictly adhere to the  steepest descent path. There is clearly an infinite number of such minimizing paths. 

The minimum energy path in computational chemistry represents one particular type of minimizing path for the minimax problem \eqref{minmaxF}, which, in addition, adheres to the gradient descent direction when moving away from the saddle point.
The  string method \cite{String2007} computes the MEP   by   evolving  the gradient descent dynamics for every points on the path until  the following   condition   is achieved:  
\begin{equation} \label{MEP-tangent} \nabla U(\varphi^*)^\perp \equiv    0. \end{equation}
 The notation  ``$\perp$'' means the projection of the vector onto the normal hyperplane at the MEP $\varphi^*$. 
In other words, the parallel condition for the tangent, $\nabla U(\varphi^*(s)) \parallel  \partial_s \varphi^*(s)$, 
holds.  
The MEP is also well known as the most probable path for the over-damped Brownian particle of \eqref{SDE},
which can be rigorously  derived by the Freidlin-Wentzell large deviation principle.
We will elaborate this point in the section of minimum action path later.

{\bf Maximum Flux Path.}~
 Berkowitz and co-workers \cite{Berkowitz1983} in 1983 showed  the  diffusive ``flux''  of  \eqref{SDE}   along a given dominant path 
is  proportional to  the line integral   $\frac{1}{\int_L  \exp(\beta U(\varphi)) \d\varphi}$, and thus designated the optimum reaction path as the minimum of  the integral 
\begin{equation} \label{Res}
  \int_L  \exp(\beta U(\varphi)) \d\varphi = \int_0^1  \exp(\beta U(\varphi(s))) \abs{\varphi'(s)} \d s. 
\end{equation}
The minimizer of \eqref{Res} is   referred to as ``minimum resistance path''  \cite{Berkowitz1983} or  ``maximum flux path''  \cite{Elber1990}.
We use  the latter name here  and call this minimizing path as  ``max-flux path'' or ``MFP'' in short.
The temperature here explicitly appears so the MFP is   temperature-dependent, but
  we should bear in mind that this proposal of MFP for representing  reaction pathways  \cite{Berkowitz1983} is justified when the energy barrier is higher than the typical magnitude of 
noise fluctuation $O(\beta^{-1})$,  since it requires  two assumptions of (i)
 stationary and constant flux and (2) ignore the reactive flux that deviates from the optimal path.
 It is noted  that this  definition of MFP can be also derived by the Transition Path Theory \cite{TPT_review,TPTICIAM}
 under the same  assumption of the existence of  a dominant thin transition tube.

The  MaxFlux algorithm  \cite{Elber1990,HuoStraub1997} applied the chain-of-states to optimize the discrete version of \eqref{Res} together with added restraints for numerical stability.
 A temperature-dependent nudged-elastic-band algorithm    \cite{TempNEB2003} used the same idea to calculate  the max-flux path too.
There are other discussion and applications related to  this variational principle of maximum flux or minimum resistance  \cite{Skeel2016,CAMERON2013137}.

{\bf Minimum Action Path.}~
The minimum action path minimizes the Freidlin-Wentzell action functional
based on the large deviation principle \cite{FW2012}. Informally, 
for the Ito stochastic differential equation
$\d X^\eps_t =b_t(X^\eps_t)\d t + \sqrt{ \epsilon} d W_t$
the Freidlin-Wentzell large deviation theory \cite{FW2012}
asymptotically estimates the probabilities  of  the solution $X^\eps_t$ within $t\in[0,T]$,
$\eps \log \mathbb{P}(X^\eps \in A )\approx -\inf_{\phi\in A}S_T[\phi]$,
at the vanishing noise limit $\epsilon\to 0$, where the action functional $S_T[\phi]:=\frac12 \int_0^T \abs{\dot{\phi}_t-b_t(\phi_t)}^2 \dt.$
 The   minimum action methods numerically solve  the minimizer of  $S_T[\phi]$ for $\phi\in AC_{a,b}([0,T];\Real^n)$. 
 The geometric minimum action method   \cite{Heymann2006}    further focuses on  the optimal time interval $T$, $\inf_T \inf_\phi S_T[\phi]$
and   minimize the following geometric action
\begin{equation} \label{gaction}
\begin{split}
\widehat{S}[\varphi] :=& \int_0^1  \abs{b(\varphi(s))}\abs{\varphi'(s)} - \inpd{b(\varphi(s))}{ \varphi'(s)} \d s 
\\
=&
2\int_L \abs{b(\varphi)}\sin^2 \frac12 \eta \d \varphi
\end{split}
\end{equation}
over all possible $\varphi\in AC_{a,b}([0,1];\Real^n)$.
Here $\eta(s)$ is the angle between the tangent $\varphi'(s)$ 
and the drift $b(\varphi(s))$.   
For the gradient system we focus on here, 
  $b(x)=-\nabla U(x)$,  so the second term 
$-\int b(\varphi(s))\cdot \varphi(s) \d s=
\int \frac{\d}{\d t} U(\varphi(s))\d s = U(\varphi_1)-U(\varphi_0)$ is independent of the path. Then we only minimize the first term
\begin{equation} \label{Ig}
    I_g[\varphi] := \int_L  |\nabla U(\varphi)| {\rm d}\varphi = \int_0^1 |\nabla U(\varphi)| |\varphi^\prime| {\rm d}s.
\end{equation}
This variational form was previously identified by Elber and Olender  \cite{olender-yet}  and termed ``scalar work''.

For the double-well potential where two neighboring  local minimum points  are separated by a saddle point
with the  lowest-energy, the minimizer of functionals \eqref{minmaxF} and \eqref{Ig} are the same: they both correspond to the MEP passing that saddle point  on the separatrix.
In this case,  the  optimal path is split into two parts.
For the uphill part of the path,
$\nabla U $ and $\varphi'$ are in the same direction and the contribution to $I_g$ is 
$U(\varphi_{sad}) -  U(\varphi_{0})$.
For  the downhill   part of the path, $\nabla U $ and $\varphi'$ are in the opposite direction and the contribution to $I_g$ is 
$  U(\varphi_{sad})-  U(\varphi_{1})$.
So the minimum  action value of $I_g$ is $2 U(\varphi_{sad})-2 U(\varphi_0)$, twice of the energy barrier.

For the more complex energy landscape with multiple meta-stable states,
the path may involve some intermediate states (the local minima besides $a$ and $b$),
then the max-flux path at the large $\beta$ limit is associated with the maximum value of the potential along the path;
while the minimum action path corresponds to the sum of energy barriers for the noise to overcome every local minima. 
Refer to  the example in Section \ref{Tw_case} for illustrations of this difference.

 \subsection{MFP  approximates the MEP at low temperature}

Based on the functional in \eqref{Res} for the temperature-dependent  max-flux path, 
we introduce \begin{equation} \label{softmax}
I_\beta [\varphi]:= \frac{1}{\beta}  \log  \int_0^1 \exp\left( \beta U(\varphi(s) ) \right)  \abs{ \varphi'(s)}\d s
\end{equation}
Minimizing  \eqref{Res} is equivalent to  minimizing the   functional $I_\beta$.
We note that 
\eqref{softmax} is the continuum version of the so-called ``LogSumExp'' function, 
a smooth approximation to the maximum function.
This     ``log-sum-exp trick''   is 
widely used in many machine learning algorithms, including the softmax function for classification. A similar form of \eqref{softmax} is also common in  the literature of large deviation theories.

The Laplace's method for integrals shows that for a twice differential function $f$, 
$$\int_0^1 \exp( \beta f(s)) \dd s \approx \sqrt{\frac{2\pi}{|f''(s_*))|}} \exp(\beta f(s_*))$$  as $\beta\to+\infty$, if $s_*\in (0,1)$ is the global maximum point of $f$.
So we have that at large $\beta$,
$I_\beta[\varphi] \approx \displaystyle \max_{0\le s \le 1} U(\varphi(s)) + O( {\beta}^{-1}).$ This is the point-wise convergence of $I_\beta \to I_\infty$
where $$I_\infty[\varphi]:=\max_{0\le s \le 1} U(\varphi(s))=\norm{U\circ \varphi}_\infty.$$
 It is obvious  that $I_\beta[\varphi] \le I_\infty[\varphi] $.
It holds  that the original minimax problem \eqref{minmaxF} for the MEP 
is approximated by 
 $
   \min_\varphi \max_{s} U(\varphi)  \approx \min_\varphi  I_\beta [\varphi]. $

\begin{thm} \label{MFP-conv-thm}
If the function $U$ is Lipschitz continuous, then  
the  functionals $I_\beta$  Gamma converges to 
the  point-wise limit $I_\infty$,
with the topology of  AC norm as $\beta\to+\infty$.
\end{thm}
\begin{proof}
The proof is based on   Proposition 5.9 in the book  \cite{MasoBook}
which states the equivalence between the Gamma convergence and the point-wise convergence,  if 
 the   so called 
equi-lower semi-continuous condition (Definition 5.8 \cite{MasoBook}) holds  at every $\varphi\in AC_{ab}$ .  Let $\beta_n$ 
be an increasing sequence satisfying $\lim_n \beta_n =+\infty$. 
The sequence $\{I_{\beta_n}: n\ge 1\}$ is  said to be {\emph{equi-lower semi-continuous}} at 
$\varphi$, if 
for every $\varepsilon>0$ there exists a neighbour $\mathcal{N}$
of $\varphi$  such that $I_{\beta_n}[\wt{\varphi}] \ge  I_{\beta_n}[\varphi]-\varepsilon$ for every  $\wt{\varphi}\in \mathcal{N}$ and every $n$.

Indeed, let $\varphi$ and $\wt{\varphi}$ be two  arbitrary paths
satisfying $\norm{\wt{\varphi}- \varphi}_\infty+\norm{\wt{\varphi}'- \varphi'}_1 \le \delta $, 
then by the Lipschitz condition of $U$ with a  Lipschitz constant $L>0$, we have
$U(\wt{\varphi}(s)) \ge U(\varphi)-  L  \delta $.
Multiplying this equality by $\beta$ and then 
taking the line integrals,
we have $I_\beta[\wt{\varphi}] - I_\beta[{\varphi}] \ge - L\delta$.
For every $\epsilon>0$, the simple choice of $\delta =\epsilon/L$ for the size of the neighbour $\mathcal{N}$
gives the desired equi-lower semi-continuous condition.
  
 \end{proof}

The Gamma convergence of the sequence $I_\beta$ as $\beta \to \infty$ 
implies the convergence of the minimum value of $I_\beta$ to 
the minimum value of $I_\infty$.
 Moreover,  any  limit of the minimizing path of $I_\beta$ is also a 
 minimizing path of the Gamma limit  $I_\infty$,  under a  compactness condition.
We can also examine this convergence issue by 
 the Euler-Lagrangian equation for minimizing ${I_\beta}[\varphi]$,
 which is   $$ \nabla^\perp  U(\varphi(s))  = \frac{1}{\beta}\kappa(s) \mathbf{n}(s)$$ 
 where $\kappa(s)$ is the curvature of $\varphi$ and 
 $\mathbf{n}(s)$ is the normal direction.
 The derivation is shown in   the  Appendix.
 As $\beta\to\infty$, this condition becomes
  the tangent condition \eqref{MEP-tangent}  exactly
  characterizing the MEP. 
  Such a connection between the MFP and  the MEP 
  serves as the theoretical foundation 
  of our numerical  pre-training strategy  
  to accelerate the MEP computation, which will be
  developed later.

\section{\label{main_method} StringNET: Numerical methods} 
The main task  of StringNET is to develop an approach  to locate the transition paths introduced previously  by using the  deep learning techniques. In this section, we will  explore  
the construction of  a few loss functions for these paths 
and compare their empirical performance. 
We start with the numerical representation of the path
by the neural networks.

\subsection{Neural network architecture}
\label{ss:NN}
 
\begin{figure*}[htbp]
\centering
 \subfigure[ {A single (fully-connected) neural network with input $s\in[0,1]$ and the output  $\varphi(s)\in\Real^d$.}]
 {
\includegraphics[width=0.6\textwidth]{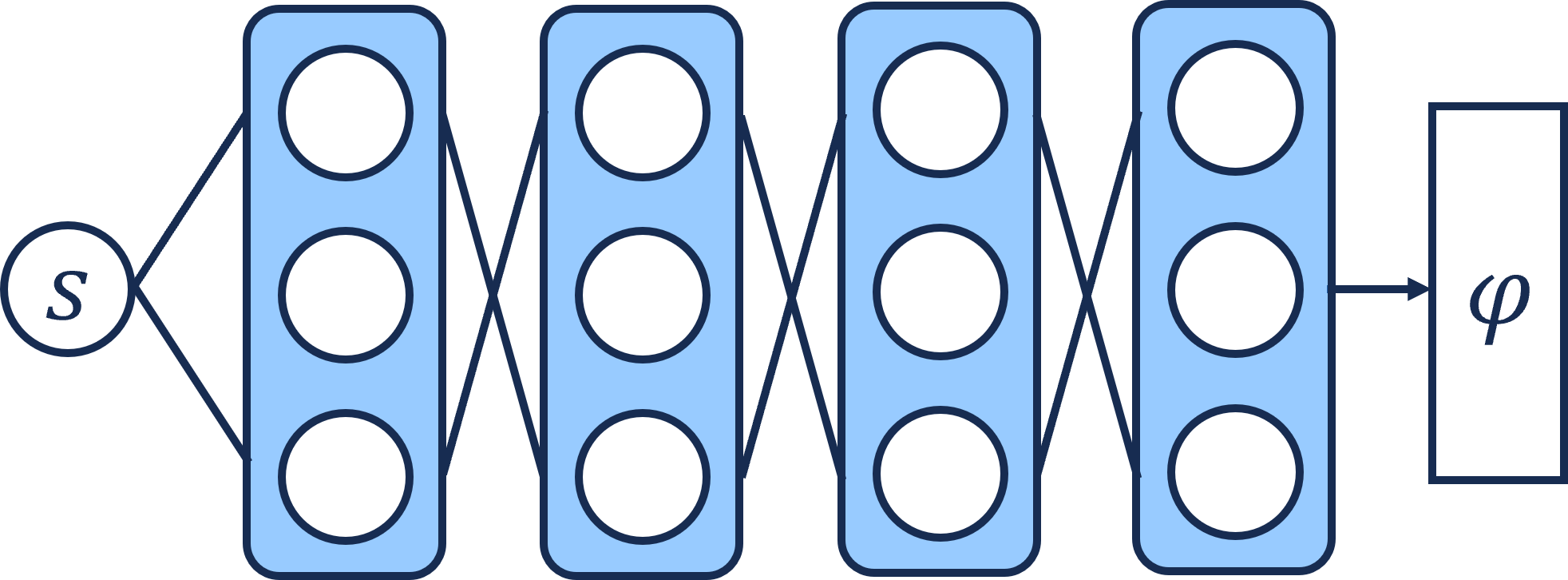}
      \label{fig:fcn}
  }
  
  \subfigure[ {$d$ separate neural networks 
for each component $s\mapsto \varphi_i(s)\in \Real^1$, $1\le i  \le d$.}]
 {
    \includegraphics[width=0.6\textwidth]{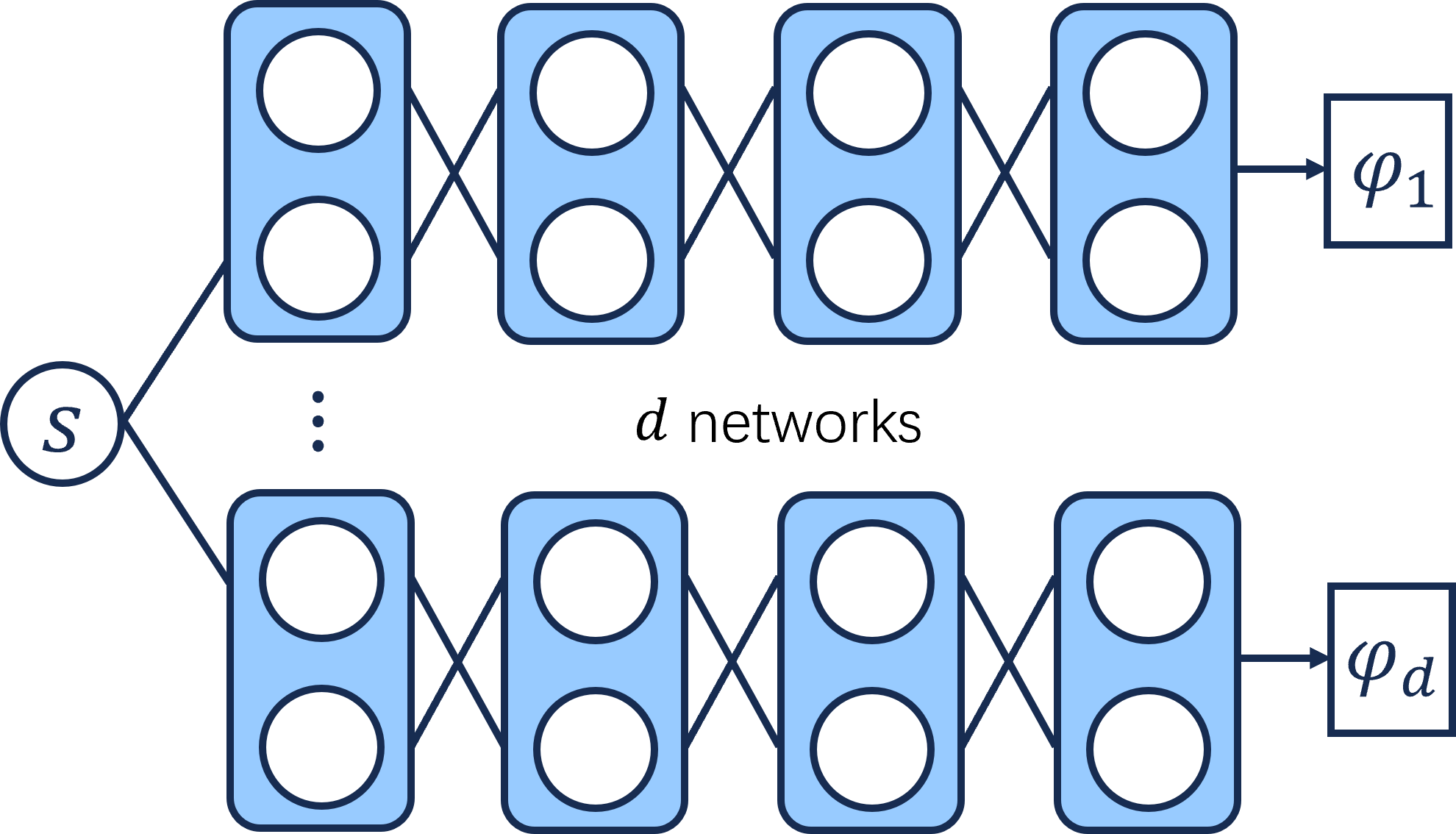}
      \label{fig:separate}
  }
 
\caption{One single neural network (a)  used in the StringNET method. 
In case of the path in the function space $\varphi(s,x)$, the input is modified as  $(s,x)$ where $x\in \Real^n$ is the spatial variable and the output becomes $\Real^1$-valued. Separate networks in (b)
for each component are neither efficient in $\Real^d$ nor applicable for PDE examples.}
 \label{fig:NN}

\end{figure*}

Since the two endpoints of the path at $s=0$ and $s=1$, are fixed
at $a$ and $b$ respectively, we use 
\begin{equation}\label{NN}
\varphi_\theta(s):=h_\theta (s)(1-s)s+ \bar{\phi}(s)
\end{equation}
where  $\bar{\phi}(s)$ is  an {\it aprior} guess   satisfying the endpoints condition, and 
$h_\theta (s)$ denotes  the  neural network function
which      maps  the arc length parameter $s\in[0,1]$ to $\Real^d$. $\theta$ refers to the generic parameters in the neural network such as the weights and bias.
Since    the initialized neural network function $h_\theta$ is very small,
the {\it aprior} guess 
$\bar{\phi}$ can be understood as the initial guess in the traditional optimization method.
For example, if the initial guess is the line segment between $a$ and $b$, then   
\begin{equation} \label{line-init}
\bar{\phi}(s)=(1-s)a +sb
\end{equation} 
In general,  the Lagrangian interpolation can be used to construct $\bar{\phi}$.
For instance, if we additionally require  $\bar{\phi}$ to pass  an intermediate state $c$ at $s=1/2$ , then   $\bar{\phi}(s)$
is quadratic:  $\bar{\phi}(s)=(1-s)(1-2s)a +s (2s-1)b + 4s(1-s)c.$

For finite dimensional problems where the path
takes value in $\mathbb{R}^d$, the path is a one-dimensional
to $d$-dimensional function, so one can either build $d$ independent neural network functions for each component as shown in Figure \ref{fig:separate} 
or  treat the path as a  
one-dimensional to $d$-dimensional function represented by  a single 
neural network  (see Figure \ref{fig:fcn} for 
an example of a fully-connected neural network(FCN) ).
We adopt the latter architecture because
it is easier to extend to infinite-dimensional spaces for scalability, where
the path takes values in function space. 
Meanwhile,  we shall show later through numerical comparisons that  the use of
one neural network  can improve the training process with much reduced cost, for the reason that the shared weight parameters  potentially unifies the learning for better consistency and efficiency. 
We use the fully-connected feed-forward networks and the sigmoid activation functions for all examples in this paper. The fine tune of the detailed architecture 
and the activation function are not further discussed.

\subsection{Loss function for arc length parametrization}
In any numerical computation of the path, it is very important to enforce
the correct parametrization of the path geometrically.
In the chain-of-state methods such as the string method and NEB,
the neighboring configurations on the path should be equidistant so that the parameter $s$ is the arc-length parameter. In our setting here,
there are no discrete points, so the arc-length parametrization is
implemented in a natural way by its definition:
 $\abs{\partial_s\varphi(s)} = const.$ 
This is equivalent to the following orthogonality condition
\begin{equation}\label{equa-arclen}
  \partial_s  \abs{\partial_s\varphi(s)}^2 = 2\partial_s \varphi(s) \cdot 
  \partial_{ss}^2\varphi(s) =0, \quad \forall s \in [0,1],
\end{equation}
and is implemented in practice by the following $L^2$ penalty function \begin{equation}\label{equal-arclength}
\begin{split}
 \ \ell_{arc}(\theta) :=&   \int_0^1
    \qty[\partial_s \varphi_\theta (s) \cdot 
  \partial_{ss}^2\varphi_\theta (s)  ]^2  \dd  s
  \\
  \approx &
  \  \frac{1}{M} \sum_{i=1}^{M} 
   \qty[\partial_s \varphi_\theta (s_i) \cdot 
  \partial_{ss}^2\varphi_\theta (s_i)  ]^2  .
  \end{split}
\end{equation}
where $s_i$ are the grid points for numerical
integration on $[0,1]$.
The derivatives for $s$ 
are computed by the automatic differentiation
of the neural network functions in    
Python packages like PyTorch or TensorFlow.
In the work of Simonnet \cite{SIMONNET2023112349}, the arc-length parametrization is imposed by minimizing the variance:
$\int |\partial_s \varphi(s)|^2   \dd s - (\int |\partial_s\varphi(s)| \dd s )^2=0$.  These two conditions are equivalent, except that the $\ell_{arc}$ in \eqref{equal-arclength} 
contains the second order derivatives (the curvature).
In  case that the path is not in $C^1$ which only occurs for non-gradient system, the other derivative calculation 
 \cite{CiCP2018-SUNZHOU} might be in need.

\subsection{\label{MFP_MEP}Loss functions for max-flux path}
 
The loss function for the max-flux path is from the variational problem \eqref{softmax}:
\begin{equation}\begin{split}
\ell_\beta(\theta) & :=I_\beta [\varphi_\theta] = \frac{1}{\beta} \log \int_0^1 \exp \left( \beta U(\varphi_\theta(s) \right) |\varphi_\theta^\prime(s) | \dd s 
    \\
   &\approx  \frac{1}{\beta} \log \left[ \frac{1}{M} \sum_{i=1}^{M}  \exp ( \beta U(\varphi_\theta (s_i)) ) |\varphi_\theta^\prime(s_i) | \right].
    \label{loss1} 
   \end{split}
 \end{equation}
This optimization problem is usually quite efficient because, 
unlike other loss functions discussed later, only the potential 
function appears in this loss function -- there is no
$\nabla U$ term in $\ell_\beta$.
The parameter $\beta$ here is the inverse temperature,
characterizing the impact of the finite size of the noise.
If one wants to approximate the MEP for the zero temperature limit based on Theorem \ref{MFP-conv-thm}, a large $\beta$ is preferred. In practice, we find that the numerical MFP at a moderate $\beta$ is  a  very good initial guess for the algorithms 
for the MEP.
In addition,   by noting that the minimizer does not change if $U$ is subtracted by a constant,
one can apply the  trick in practice by replacing the
absolute $U$ by the difference $U-U(a)$ when necessary.

\subsection{Loss function based on least squared  tangent condition  for minimum energy path} \label{ssecD}

The minimax formulation for the MEP in \eqref{minmaxF}
cannot be directly applied in  computations due to the maximum norm. The approximate by the MFP with large $\beta$ 
is hindered by the truncation of $\beta$ from the infinity. 
 Alternatively,  we consider to  solve the  first order condition
\eqref{MEP-tangent} for the MEP.
Let  $\eta(s)$ denote the angle between the gradient $\nabla U(\varphi (s))$ and the tangent $\partial_s \varphi(s)$.
Equation \eqref{MEP-tangent} implies that  $\eta(s)=0$ or $\pi$,  i.e., 
$   \sin \eta (s)\equiv 0$.
This condition  can be addressed  by   the following minimization problem for the  mean squared sine value, with   an arbitrary weigh function $w(s)>0$,
\begin{align}
\label{eq:loss3}
    \ell_\parallel[\theta]:=&  \int_0^1 
   ( \sin \eta(s))^2 ~  {w}(s) {\d}s 
     \nonumber 
     \\
    = &\int_0^1 \Big(1-\frac{\inpd{\nabla U (\varphi_\theta(s))}{\partial_s \varphi_\theta(s)}^2}
    { \abs{  \nabla U (\varphi_\theta(s)  } ^2 
    \abs{  \partial_s \varphi_\theta(s))    } ^2} \Big) \, w(s) {\d}s, \nonumber 
    \\
    \approx & \frac{1}{M}\sum_{i=1}^{M} \Big( 1-
    \frac{\inpd{\nabla U (\varphi_\theta(s_i))}{\partial_s \varphi_\theta(s_i)}^2}{|\nabla U(\varphi_\theta(s_i))|^2 |\partial_s \varphi_\theta(s_i)|^2} \Big) w(s_i),
\end{align}
For simplicity,  we simply let $w$ be constant. It is noted that this method only addresses the necessary condition \eqref{MEP-tangent}. In theory, any heteroclinic orbit connecting $a$ and $b$ and passing through any critical point on the boundary also satisfies this condition \eqref{MEP-tangent}. More importantly, unlike the string method \cite{String2002} where the images on the path can follow the gradient descent independently, the optimization of $\ell_\parallel$ is much less intuitive due to the neural network parametrization.  We shall see that the numerical minimization of $\ell_\parallel$ is more challenging, requiring a high-quality initial condition. In addition, there is a numerical challenge near the critical value $s$ where $\nabla U(\varphi(s)) = 0$.

One remark is that if $w$ is chosen properly instead of the the , then we have a new version of the loss funciton, $\int_0^1 \Big({ \abs{  \nabla U (\varphi_\theta(s)  } ^2 
    \abs{  \partial_s \varphi_\theta(s))    } ^2}- {\inpd{\nabla U (\varphi_\theta(s))}{\partial_s \varphi_\theta(s)}^2}
     \Big)   {\rm d}s$, which is similar to the geometric action loss in Equation \eqref{gaction} except the squars.

\subsection{Loss function for minimum action path in gradient systems} \label{ssecE}
By the geometrical functional \eqref{Ig}. Therefore, we consider the following loss function to find the mininum action
path for the gradient system:
\begin{align}
\label{loss-mam}
    l_g(\theta) : = I_g[\varphi_\theta(s)]=  \int_0^1 
    \abs{\nabla U(\varphi(s))} \abs{\partial_s \varphi (s)} {\rm d}s  
\\
    \approx \frac{1}{M}\sum_{i=1}^{M} 
    |\nabla U (\varphi_\theta(s_i)|  \abs{ \partial_s \varphi_\theta (s_i)}.
\end{align}
The optimization can be handled by autodifferentiation even for the non-smooth 
norm function $\abs{\cdot}$.  
One can also use an iterative (re-)weighted scheme to 
minimize the squared error with a frozen weight 
$  \int_0^1 \abs{\nabla U(\varphi)}^2  \abs{\varphi^\prime}^2 w_k(s) \d s $  at each iteration $k$ with a temporally fixed weight $w_k(s) \propto 1/ \abs{\nabla U(\varphi_{k-1}(s))} \abs{\partial_s \varphi_{k-1} (s)}$.
This loss function involves the gradient force
$\nabla U$, so the gradient descent optimization 
needs the Hessian function of $U$.  Consequently,  the training efficiency of $\ell_g$ is generally
  less than that of the loss $\ell_\beta$ for max-flux path.

\subsection{The sum of 
  loss functions with weights tuning, and pre-training for MEP}
    
To summarize, for the convenience of practical computation, we present a general form by summing each loss function together:

   \begin{equation}
     \min_\theta J(\varphi_\theta)
 \end{equation}
where
\begin{equation}\label{MEP_loss_new}
     J(\varphi_\theta):= \alpha_1 \ell_{\beta} + \alpha_2  \ell_{arc} + \alpha_3 \ell_\parallel + \alpha_4 \ell_g.
 \end{equation}
By setting various values of weights $\alpha_i$, we can achieve different goals. 
The  weight $\alpha_2$ for  arc-length parametrization loss $\ell_{arc}$  in \eqref{equal-arclength}  is always  necessary for any path calculation.
 It is recommended that this weight not be set too large initially, since the minimizers of 
$\ell_{arc}$ alone are not unique and  tend, empirically,  to form a straight line.
To assess the quality of the arc-length parametrization, 
we consider the indicator 
$\gamma=\frac{\max_s \abs{\partial_s \varphi} } {\min_s \abs{\partial_s \varphi}}$.
 If $\gamma$ is larger than a threshold (for example, 5),
we increase the weight by $\alpha_2 \to \gamma \alpha_2$.

$\ell_\beta$ 
 is for temperature-dependent max-flux path (MFP), 
and $\ell_\parallel$ and $\ell_g$ are   for the MEP and MAP, respectively.
We can also schedule their weights during the training. 
We shall show later that the computation of the MEP based on $\ell_\parallel$ or $\ell_g$
is    generally    challenging, mainly due to the initial guess of the path (a straightforward line most of time).
Therefore we propose to additionally turn on the weight $\alpha_1$ for the loss $\ell_\beta$ with a moderate $\beta$ at least in the early stage of the training.  Extensive numerical tests show
this pre-training technique   using $\ell_\beta$ significantly boosts the optimization efficiency.
 In practice, we also find that if  for the pre-training purpose only,   one drops the arc-length $\abs{\varphi'(s)}$ in the 
 integral of  $\ell_\beta$, and uses instead 
\begin{equation}\begin{split}
\wt{\ell}_\beta(\theta) & :=  \frac{1}{\beta} \log \int_0^1 \exp \left( \beta U(\varphi_\theta(s) \right)  \dd s 
     \approx  \frac{1}{\beta} \log \left[ \frac{1}{M} \sum_{i=1}^{M}  \exp ( \beta U(\varphi_\theta (s_i)) )   \right],
    \label{loss11} 
   \end{split}
 \end{equation}
then the optimization efficiency is further improved. 
 When we refer to the pre-training later, we mean to use the above simplified version of $ {\ell_\beta}$.  When we compute the max-flux path for a  $\beta$,  the  true loss \eqref{loss1} 
 in the form of   line integral is applied.

  \subsection{Generalization to energy functional
  on Hilbert space}
  \label{ssec:PDE}
  All discussions above on a potential energy function  $U: \Real^d \to \Real$ can be readily 
  generalized to an  energy functional
  $E(u)$ for $u$ in a  Hilbert space $H$, for example,   the Ginzburg-Landau free  energy functional 
  \begin{equation} \label{GL-ef}
  E(u)=\int_\Omega   \dfrac12 \kappa^2  \abs{\nabla u}^2 + f(u(x)) \dx 
  \end{equation}
  for a domain $\Omega\subset\Real^n$ and a double-well   $f(u)=\frac14 (1-u^2)^2$.  $\kappa$ is a small  positive parameter related to the thickness of the interface between phases.
   The path and its neural network realisation are   then written as $\varphi(s,x)$ and $\varphi_\theta(s,x)$:  $ [0,1] \times \Real^n  \to \Real$  by concatenating  the arc-length parameter $s$ and the spatial variable $x$.
  
 The loss function  for the max-flux path, Equation \eqref{loss1}, now 
 takes the form 
\begin{equation}\begin{split}
\ell_\beta(\theta) &  = \frac{1}{\beta} \log \int_0^1 \exp \left( \beta E(\varphi_\theta(s,\cdot) \right) \norm{ \partial_s \varphi_\theta (s,\cdot)}_{L^2(\Omega)}   \dd s 
    \\
   &\approx  \frac{1}{\beta} \log \left[ \frac{1}{M} \sum_{i=1}^{M}  \exp \qty( \beta E(\varphi_\theta (s_i)) ) \norm{ \partial_s \varphi_\theta (s_i,\cdot)}_{L^2(\Omega)}  \right].
    \label{loss1-pde} 
   \end{split}
 \end{equation}
  where $E(\varphi_\theta(s_i,\cdot))$ is   approximated by 
  $M_i$ sample average in $\Omega$,   $$E(\varphi_\theta(s_i,\cdot))\approx \dfrac{1}{M_i} \sum_{j=1}^{M_i} 
  \qty[ 
  \dfrac12 \kappa^2 
 \abs{ \nabla_x \varphi_\theta(s_i,x_{i,j})}^2 + f (\varphi_\theta(s_i,x_{i,j})) 
 ]$$ where  $x_{i,j}$ are uniformly drawn inside the spatial domain $\Omega$.
  Likewise,  we approximate the spatial $L^2$ norm of the tangent  $\norm{ \partial_s \varphi_\theta (s_i,\cdot)}_{L^2(\Omega)}\approx  
  \sqrt{ \dfrac{1}{M_i} \sum_{j=1}^{M_i} (\partial_s \varphi_\theta(s_i,  x_{i,j}))^2 }$.
  For simplicity, we only use the same distribution of $x$ at every $s_i$, so that 
we can independently  sample the spatial points $x_{i,j}=x_{j}$ and use
the same set of $\{x_j\}$ for all $s_i$, with the same $M_i\equiv M$.
 Numerous adaptive schemes incorporating importance sampling  of  $x_{i,j}$
  have the potential to enhance statistical accuracy of $E$. However, this paper will not delve further into this particular topic.  
  
  The loss function for the minimum action path 
  in Equation \eqref{loss-mam}  is also defined similarly.
 For example, in the $L^2$ Hilbert space,  the gradient norm $\abs{\nabla U}$ is replaced now by the $L^2$ norm of 
 the first variation  of $E$:
\begin{align*} 
&~~\norm{\frac{\delta E}{\delta u} (\varphi_\theta(s_i,
 \cdot))}_{L^2(\Omega)} 
 =\norm{\kappa^2 \Delta_x \varphi_\theta(s_i,\cdot)-f'(\varphi_\theta(s_i, \cdot)}_{L^2(\Omega)}
 \\
& \approx \sqrt{ \dfrac{1}{M_i} \sum_{j=1}^{M_i} \qty(\kappa^2 \Delta_x \varphi_\theta(s_i,x_{i,j})-f'(\varphi_\theta(s_i, x_{i,j})) )^2 }.
 \end{align*}
 
 We remark that the action in Equation \eqref{loss-mam},  $$\int_0^1 
 \norm{\frac{\delta E}{\delta u} (\varphi_\theta(s_i,
 \cdot))}_{L^2} 
\norm{  \partial_s \varphi_\theta(s_i,
 \cdot))}_{L^2} 
  \dd s  $$  takes the form 
  $$ \e_{S} \qty[\sqrt{ \e_X  g(S,X) }  \sqrt{ \e_X  h(S,X) } ] $$
  for two functions $g(s,x)$ and $h(s,x)$,
  which can not be written in  the simple expectation
  form of   
  $\e_{S,X} c(S,X)\approx \frac{1}{M}\sum_i c(s_i,x_i)$ for some function $c$.
Even though the above is bounded by 
$\sqrt{ \e_S  \e_X  A(S,X)    \e_X  B(S,X) }$ by Jensen's inequality, 
to minimize this upper bound only \cite{SIMONNET2023112349} is  not equivalent to 
the original minimization problem.

\section{\label{sec:NT}Numerical Tests}

In this section, we explore the numerical performance 
  by  testing the  StringNET  over several numerical examples.  We discuss important  empirical observations and numerical techniques for accelerations too.


\subsection{Double-well   potential}\label{dw_case}

\begin{figure}[htbp]
\centering
\includegraphics[width=.8\textwidth, height=0.6\textwidth]{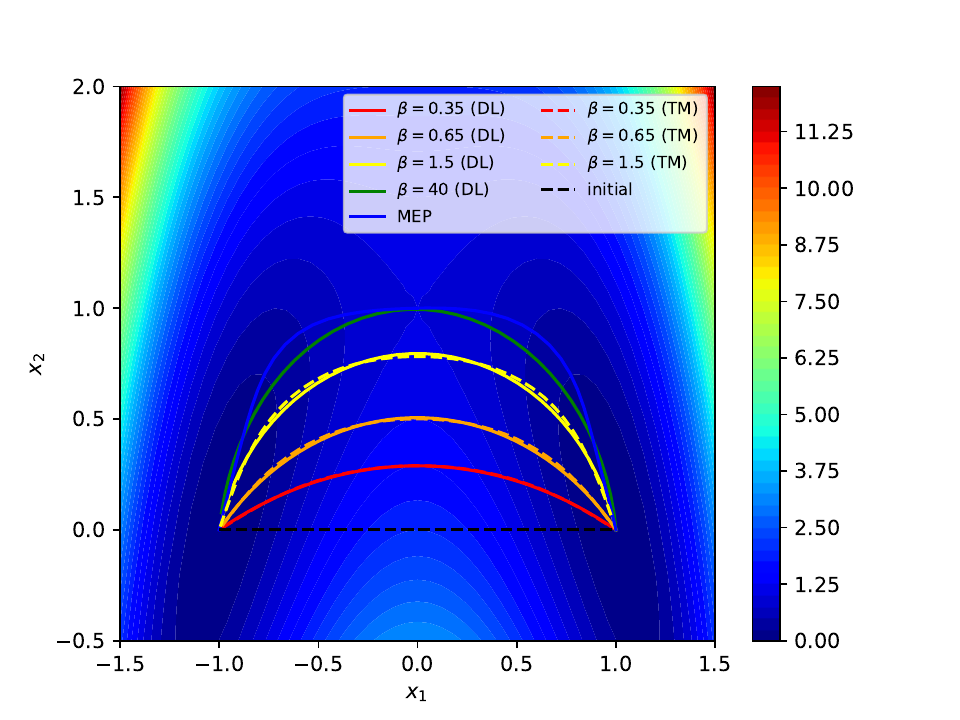}
 \caption{The initial path is the straight line 
 between two minimizers 
 $(\pm 1, 0)$ (dashed black).
 The MEP is shown in blue.
 The max-flux paths at different 
 $\beta=0.35, 0.65, 1.5, 40$ are computed by the neural network function (DL) and the traditional chain-of-states method (TM).   The latter method (TM)  
 with the MATLAB subroutine \textsf{fmincon} fails to converge
 at  $\beta=40$.  }     
 \label{MFP_dw}
 \end{figure}

The first numerical example is   the double well potential
$$U(x_1,x_2)=(x_1^2-1)^2+(x_1^2+x_2-1)^2.$$
There are two local minima $m_1 = (1,0)$ and $m_2 = (-1,0)$ and one saddle point $x_s = (0,1)$. The contour map of the problem is shown in Figure \ref{MFP_dw}.
We use this simple model to examine the neural network-based method and the different losses in practical computations.

The neural network function $h_\theta$ in \eqref{NN} is realized as a 5-layer fully connected neural network with the neuron numbers  $[1,16,32,32,16,2]$ at   each layer, respectively. 
The activation function is  the ``sigmoid''  function.
The    batch size in the ADAM optimizer  is $M=500$. 
The learning rate in the ADAM is 
$10^{-4}$.
 
We  first validate the numerical  results of the  MFP from the neural network method
with varying  $\beta$ by  setting 
 $\alpha_1=10, \alpha_2=1$ and   turning  off other  MEP losses 
($\alpha_3=\alpha_4=0$) in the total loss
 \eqref{MEP_loss_new}.
 We compare  it with the  traditional optimization in the chain-of-state methods
 by discretizing  the path. 
 Both methods 
 start from the same initial path of a straight line connecting local minima. 
The results   in Figure \ref{MFP_dw} 
show    good agreement  for three tested  $\beta$.
  
We also observe  in Figure \ref{MFP_dw} that the  MFP  tends to converge towards  the MEP as $\beta$   
 increases.  
 But at the large $\beta$ regime,  it is noteworthy that the optimization approach based on the traditional chain-of-state discretisation      \cite{Elber1990,TempNEB2003}
 exhibits instability and divergence.
  In contrast, our neural network-based variational method can effortlessly manage   $\beta$ as high as $40$.  This proficiency enables the achievement of highly precise optimal solutions for the minimax problem (see Equation \eqref{minmaxF}),  as the numerical path at $\beta=40$  
  captures the saddle point   $(0,1)$ very well.

\begin{figure}[htbp]
\centering
 	\begin{minipage}[b]{\linewidth}
        \centering
    \includegraphics[width=.8\textwidth]{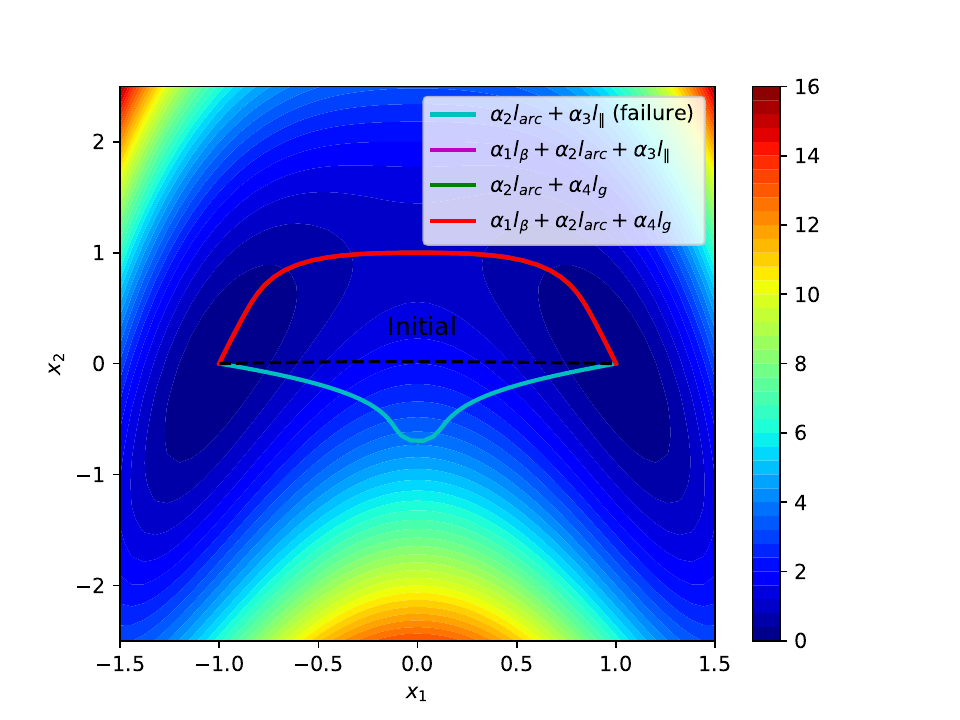}
    \end{minipage}
\caption{Comparison of numerical results for the MEP in  the double well example.
The straight line in dash is the initial. 
The neural network method
of minimizing the loss $l_\parallel$ of mean square of tangent condition (``$\alpha_2\ell_{arc}+\alpha_3\ell_\parallel$'') fails  in this simple example.  The other three results have the same numerical path.
The minimum action method for $l_g$ 
( ``$\alpha_2\ell_{arc}+\alpha_4\ell_g$'')
 works well to find the MEP.
After adding the $\ell_\beta$ loss in the early stage,  the method based on 
$l_\parallel$  can find the correct path too.
Here 
$ \alpha_2=1, \alpha_3=\alpha_4=10$ are used and 
 $\alpha_1$,  the weight for $\ell_\beta$, 
is set to drop from $10$ to $0$ after $2\times 10^4$ optimization steps. 
 }\label{fig:path_compare_dw}
\end{figure}

\begin{figure*}[htbp]
\centering

\includegraphics[width=.32\textwidth]{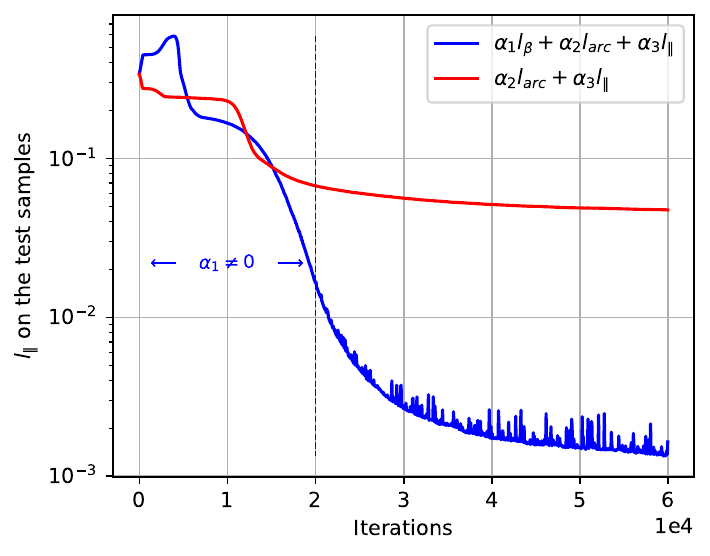}    \includegraphics[width=.32\textwidth]{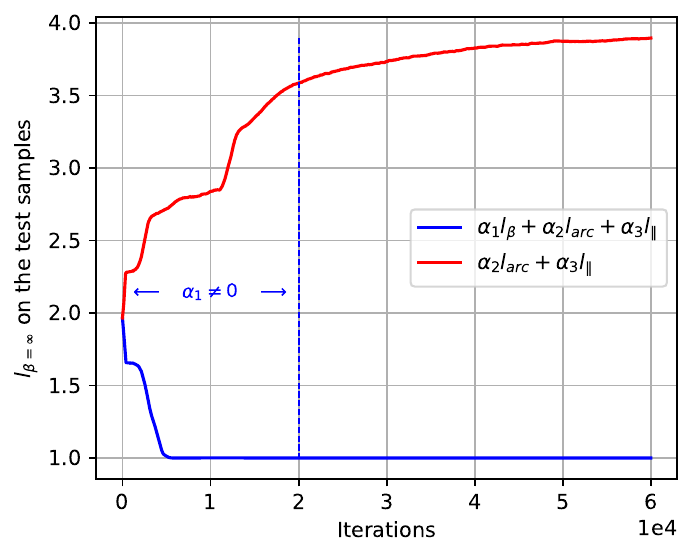}  \includegraphics[width=.32\linewidth]{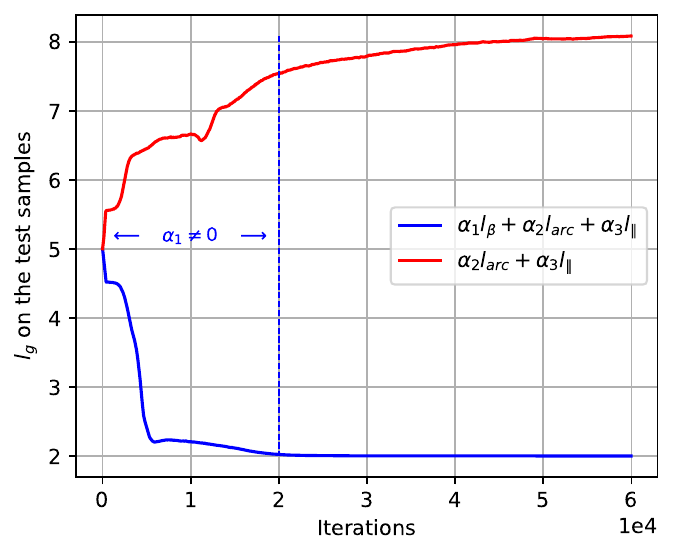}
\caption{ The loss evolution 
during the optimization iterations
by using  only   $l_\parallel$ loss
and using the combined $\ell_\beta +   l_\parallel$ loss
(See Figure \ref{fig:path_compare_dw}).
 The vertical axes in three panels   are the loss $l_\parallel$ (the optimal value is $0$),    $l_{\beta=\infty}$ (the optimal value is $1$) 
and the action $l_g$ (the optimal value is $2$) respectively,
as the three measurements of the true errors
computed  on the test  samples.} 
\label{fig:loss-dw}
\end{figure*}

We  explore the numerical 
performance of computing the MEP by 
minimizing  the loss  function  $l_\parallel$ in  \eqref{eq:loss3} 
or the action loss function  $l_g$ in \eqref{loss-mam}.
We first  find that    even for this easy test, the  minimization of $l_\parallel$ in  \eqref{eq:loss3} 
 fails to deliver the desirable result. 
  Figure  \ref{fig:path_compare_dw} compares the results from only using the loss $l_\parallel$ or only using the action loss $l_g$,   from the same straight line initial guess.
The numerical minimizing path  of  $l_\parallel$    aligns along the gradient 
direction, but carries a huge error  near the separatrix, so the optimization of $l_\parallel$
is easily trapped in spurious  local minima.
 After adding the max-flux path loss to the $l_\parallel$ loss,  one can find the right MEP again. 
 The plot of the losses during the optimization process in  Figure \ref{fig:loss-dw} confirms the
 infeasibility of minimizing $l_\parallel$ only.
 As a summary from  this example, we have the   observations  below.
  \begin{itemize}
  \item  The neural-network based variational method can manage the calculation of  the
  max-flux path in the regime of larger $\beta$ than the 
  optimization for the chain-of-states discretization.
  \item  The numerical max-flux path  can approximate  the saddle point very accurately.
  \item The   loss $l_\parallel$ fails to find the MEP and is not recommended to use alone.
    The inclusion of the MFP loss $\ell_\beta$ with a finite $\beta$  can alleviate this problem.
   \end{itemize}

\subsection{\label{Tw_case}Three-well potential: comparison of max-flux path and min action path}

The following example has multiple optimal paths and serves a good test example too.
$$
\begin{aligned}
U(x_1, x_2) &=  3 \exp \left(-x_1^2- \left( x_2-\frac{1}{3}\right)^2\right)\\& -  3 \exp \left(-x_1^2- \left(x_2-\frac{5}{3}\right)^2  \right)\\
&-5 \exp \left(-(x_1-1)^2-x_2^2\right) 
 -5 \exp \left(-(x_1+1)^2-x_2^2\right).   
\end{aligned}
$$
This potential has three local minimum points located at 
  $m_1 = (-1.1337,-0.03864)$, 
     $m_2=(1.1337,-0.03864)$,
        $m_3=(0,1.7567)$;
 and    three saddle points at 
  $s_1 = (0,-0.37157)$,
 $s_2 =(-0.69105,1.1204)$,
  $s_3 = (0.69105,1.1204)$,
 and one maximum point  $    (0,0.51824).$ The   potential function values at these critical points are also listed below. 
 
\begin{center}
    \begin{tabular}{|c|c|c|c|c|c|}
 
  \hline critical points &     $m_1 (m_2)$ & $m_3$  & $s_1$  & $s_2 (s_3)$ & $\max$ 
   \\  \hline 
  $U$ &    $-4.279$     &  $-2.748$ &  $-1.426$& $-1.756$ & $-0.715$\\
   \hline \end{tabular}
\end{center}


\begin{figure*}[htbp]
\centering
\subfigure[]
{
 	\begin{minipage}[b]{0.48\linewidth}
        \centering
        \label{MFP_3holes_init}
    \includegraphics[width=\linewidth]{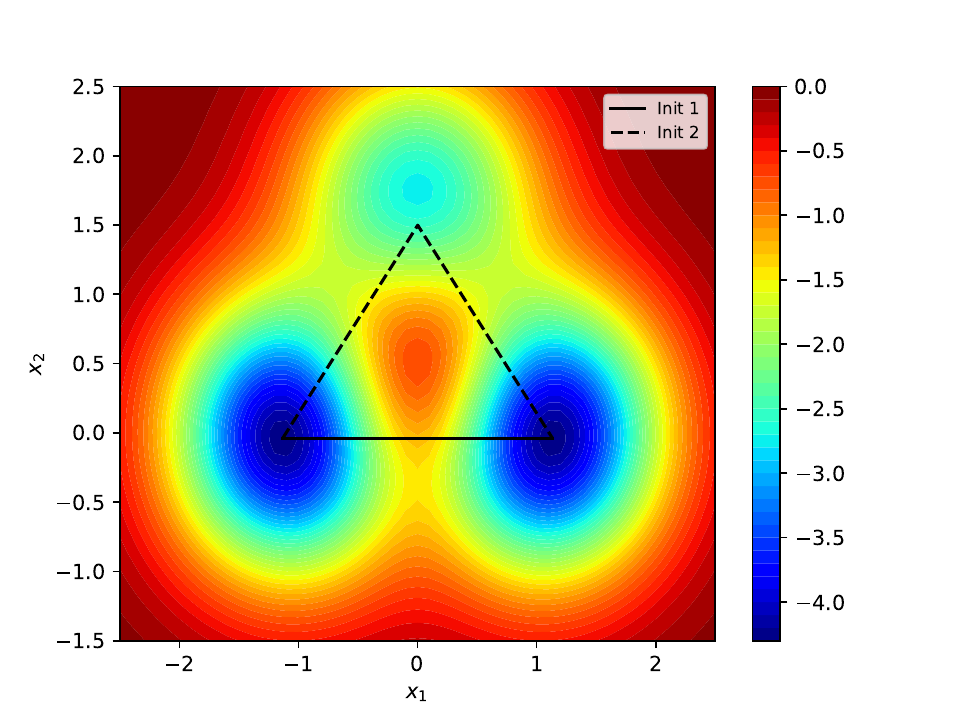}
    \end{minipage}
}
\subfigure[]
{
 	\begin{minipage}[b]{.48\linewidth}
        \centering
        \label{MFP_3holes_DL}
    \includegraphics[width=\linewidth]{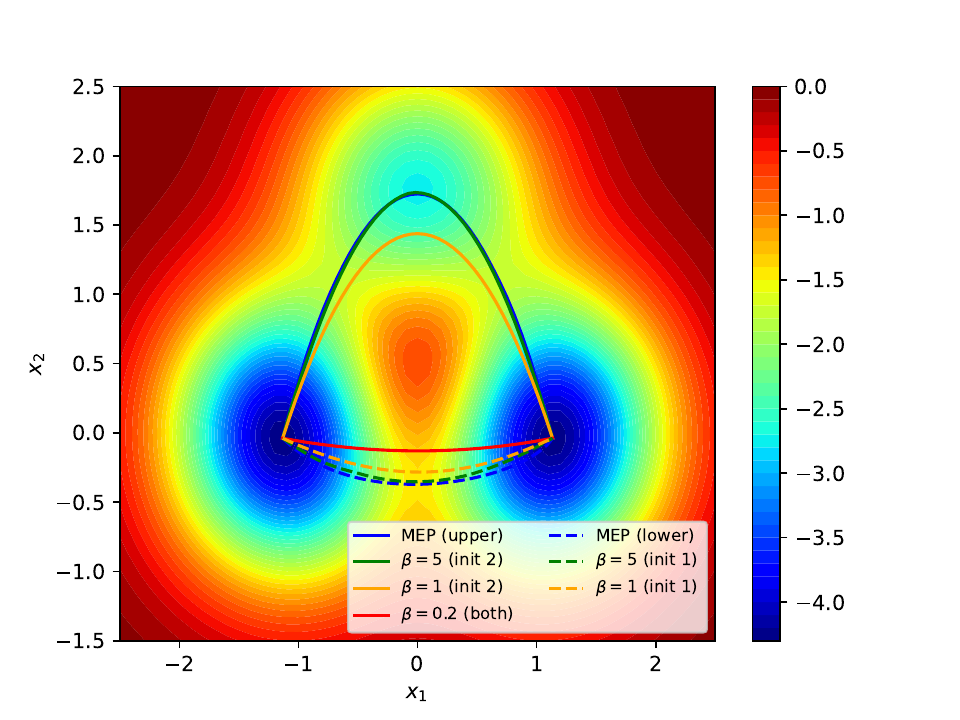}
    \end{minipage}
    }
    
\caption{ The numerical max-flux paths for various values of  $\beta$ tested from two initial paths. 
(a): two initial paths; (b):  the max-flux path and the MEP, respectively.
For $\beta=0.2$, the numerical path follows the lower branch regardless of which initial guesses in (a).
For $\beta=5$, the max-flux paths is not discernible from the upper-branch MEP.
}\label{MFP_3holes}
\end{figure*}

\begin{figure}[htbp!]
\centering   \includegraphics[width=.68\textwidth]{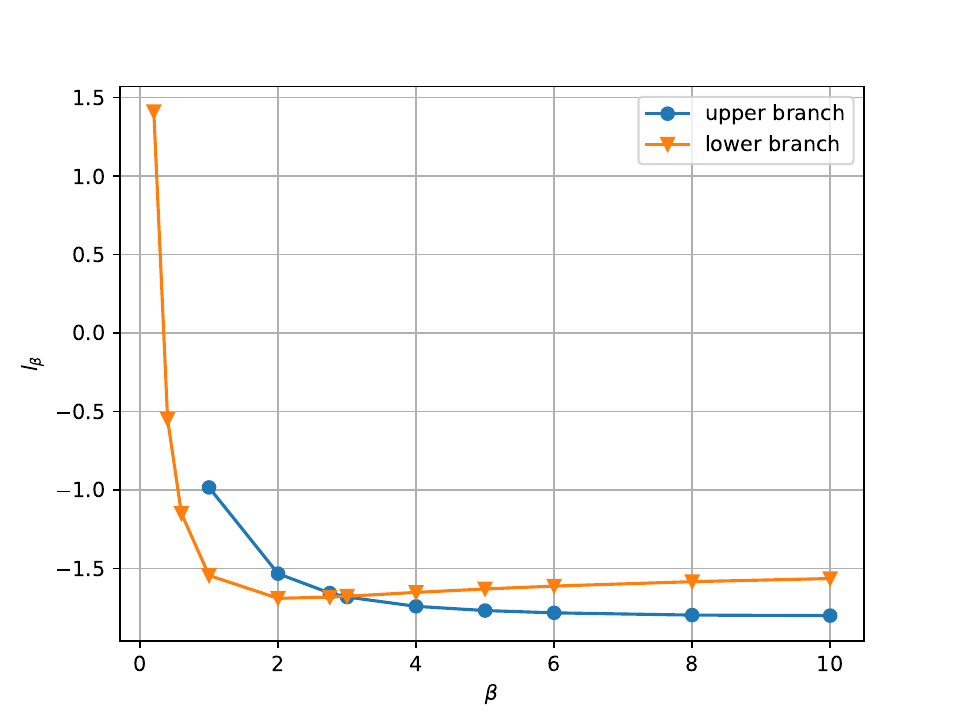}
 \caption{ Optimal values of the   objective function $\ell_\beta$  defined in Equation \eqref{softmax},
   for  the numerically determined  max-flux paths 
  at varying values of  $\beta$ using the StringNET method.
The parameters are set to     $\alpha_1=10, \alpha_2=10^{-4}, \alpha=\alpha_4=0$ 
as specified  in   \eqref{MEP_loss_new}. 
 For each selected $\beta$,   two minimizing paths are found  -- the upper branch and lower branches -- except when 
  $\beta$ is exceptionally  small (  less than approximately $0.65$), where    only  the  lower branch is found.  
  For large $\beta$ values
  (greater than approximately  $2.6$ ), tthe upper branch transitions from being a local minimizer to the global minimizer.
  As $\beta$ approaches infinity, the loss $l_g$ reaches $U(s_2)=-1.756$ for the upper branch
   and $U(s_1)=-1.426$ for the lower branch, respectively 
   }     
 \label{MFP_3holes_beta}
 \end{figure}

%

What is special about this example is that there 
are two  optimal paths of
  MEP between $m_1$ and $m_2$, due to multiple saddle points.  One path 
 follows the upper branch
via $s_2 \to m_3 \to s_3$
and the other follows the lower branch route via $s_1$ directly.

For the max-flux path, by Theorem \ref{MFP-conv-thm},
 the limit  $\lim_{\beta\to\infty}U_\beta[\varphi]$ is 
$\max_{0\le s \le 1} U(\varphi(s))$. For the upper branch path, this value  is $U(s_2)=U(s_3)$; 
and for the lower branch, it is $U(s_1)$. 
Since $U(s_2) < U(s_1)$, then the global   max-flux path at $\beta\to \infty$ is the MEP path in  the upper branch. 
The   lower branch is the local minimizing  path.
For the small $\beta\to 0$, the minimizing path is the straight line segment between $m_1$ and $m_2$, closer
to the lower branch.  Figure \ref{MFP_3holes_beta} shows the value of $\min_\varphi U_\beta[\varphi]$ in Equation \eqref{softmax}
associated with two minimizing paths (upper branch v.s. lower branch) for a sequence of increasing $\beta$, computed by the StringNET method
with two initial paths, respectively
(shown in Figure \ref{MFP_3holes_init}).
This figure validates the switch of global minimizer for the max-flux path as $\beta$ increases. 
Figure \ref{MFP_3holes_DL} plots the profiles of 
these paths at different $\beta$ and the two minimum action paths.
We also note that the traditional optimization method for chain-of-states converges  only for $\beta \leq 2.4$, while
the neural network based optimization method works well at   large values of $\beta$.

For the minimum action path,  the half  of the minimal  $l_g$ or $I_g$ in 
Equation \eqref{loss-mam} is the sum of energy barriers.
So, for the lower branch , this value is 
simply $U(s_1) - U(m_1)=2.853$;
for the upper branch, 
the total action is $3.515$, which is  the sum of two barriers   $U(s_2)-U(m_1)=2.523$
and $U(s_3)-U(m_3)=0.992$. So the lower branch MEP is the global min action path. 
%
%



\subsection{Alanine Dipeptide}

\begin{figure}[htbp]
\centering
    \includegraphics[width=.8\linewidth]{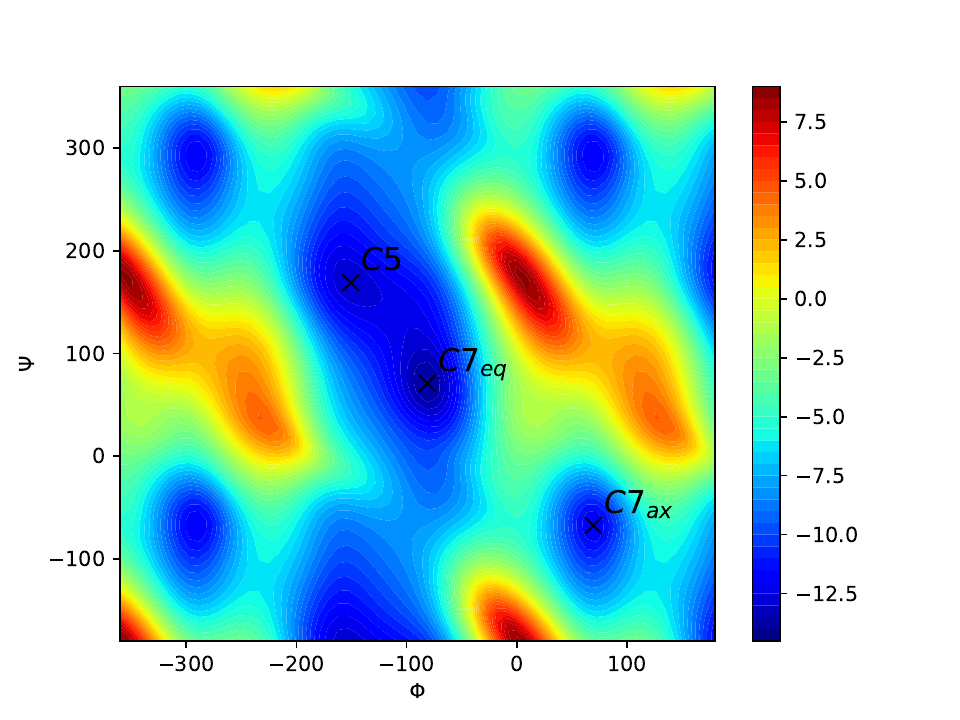}
    \caption{Free energy function $U$ in terms of two  dihedral angles, $\Phi$ and $\Psi$, for alanine dipeptide example.  To show that this energy landscape is periodic on a torus, the contour within one and half periodic box 
    ($[-360^{\circ}, 180^{\circ}]\times [-180^{\circ},360^{\circ}]$) are plotted.
  So there are four $C7_{ax}$ symmetric images shown above. }  \label{fig:ad_surface}
\end{figure}

\begin{figure*}[htbp]
\centering
       \includegraphics[width=.32\linewidth]{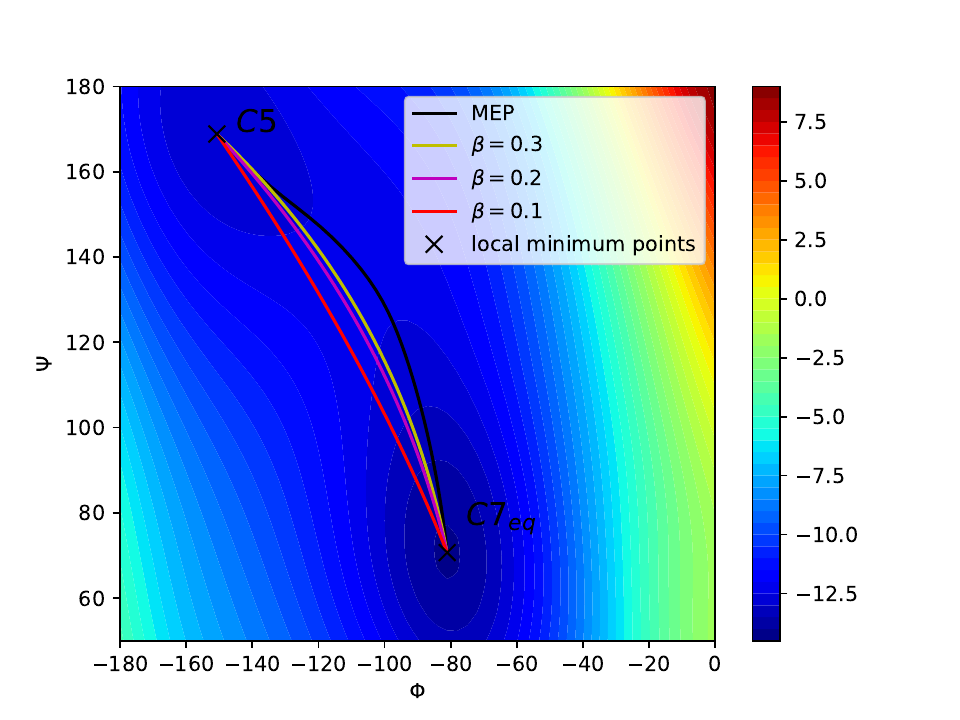}
        \includegraphics[width= .32\linewidth]{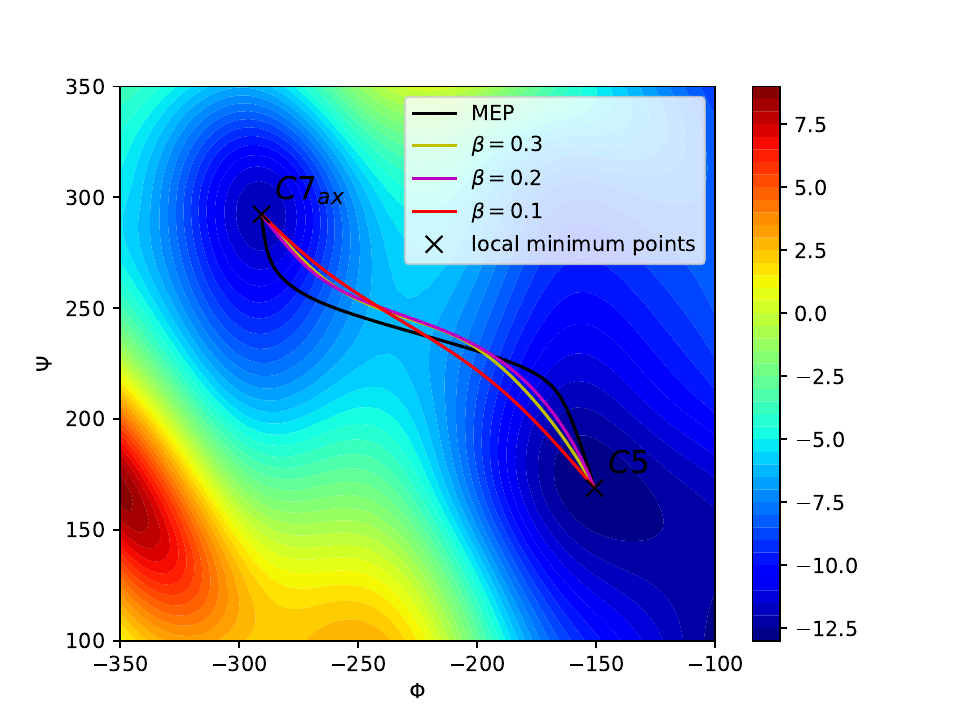}
    \includegraphics[width= .32\linewidth]{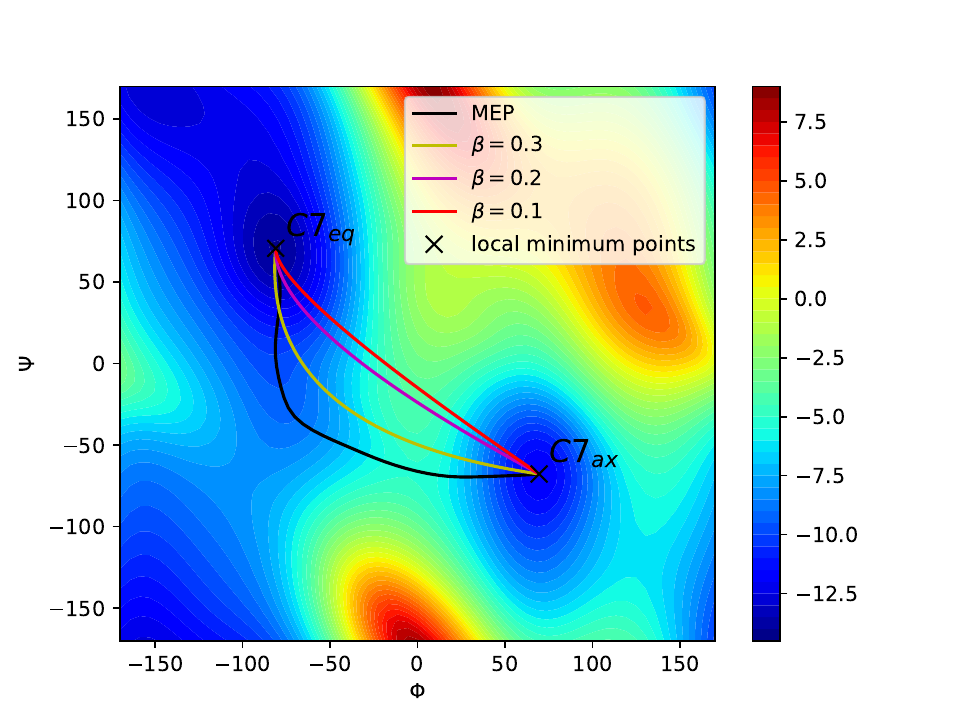}
    \caption{Max-flux paths between $C5$ and $C7_{eq}$, $C5$ and $C7_{ax}$,
   $C7_{eq}$ and $C7_{ax}$, respectively. The MEP are also shown.}  \label{fig:beta_c7c7eq}
\end{figure*}

We next show the numerical paths for  the conformation change of 
alanine dipeptide \cite{1990Alanine-dipepitide} in vacuum, a 22-dimensional molecular dynamic model  whose collective variables are two torsion angles  $\Phi$ and $\Psi$.
Here, we study the isomerization process of the alanine dipeptide in vacuum at T = 300K. The isomerization of alanine dipeptide has been the subject of several theoretical and computational studies, therefore it serves as a good benchmark problem for the proposed method \cite{E2005242}.

The free energy function $U$ for this example is defined on two  dihedral angles, $\Phi$ and $\Psi$,   in degrees. 
The expression of this function is computed via the Gaussian Process Regression (GPR)
based on the molecular dynamics simulation data at the $35\times 35$ grid in $[-180^{\circ},180^{\circ}]^2$, after 
  transforming to the  coordinate $(\sin(\Phi),\cos(\Phi),\sin(\Psi),$$\cos(\Psi))$
  to ensuring the periodicity rigorously.
The  fitted function  $U$ (shown in Figure \ref{fig:ad_surface}) is then used for our path calculation.  

 There are three local minimum points named $C_{5}, C_{7eq}$ and $C_{7ax}$, which correspond  to different isomers of alanine dipeptide. 
In Figure \ref{fig:beta_c7c7eq}, we show the max-flux paths between these three meta-stable state
computed from neural networks based on $\ell_\beta$, 
at different different $\beta$. The MEP is also drawn for reference, which is computed by minimizing $\ell_g$.

\subsection{$d-$dimensional Muller Potential}
\label{ss:MP}
\begin{figure}[htbp!]
\centering
 	\begin{minipage}[b]{\linewidth}
        \centering
    \includegraphics[width=.8\linewidth]{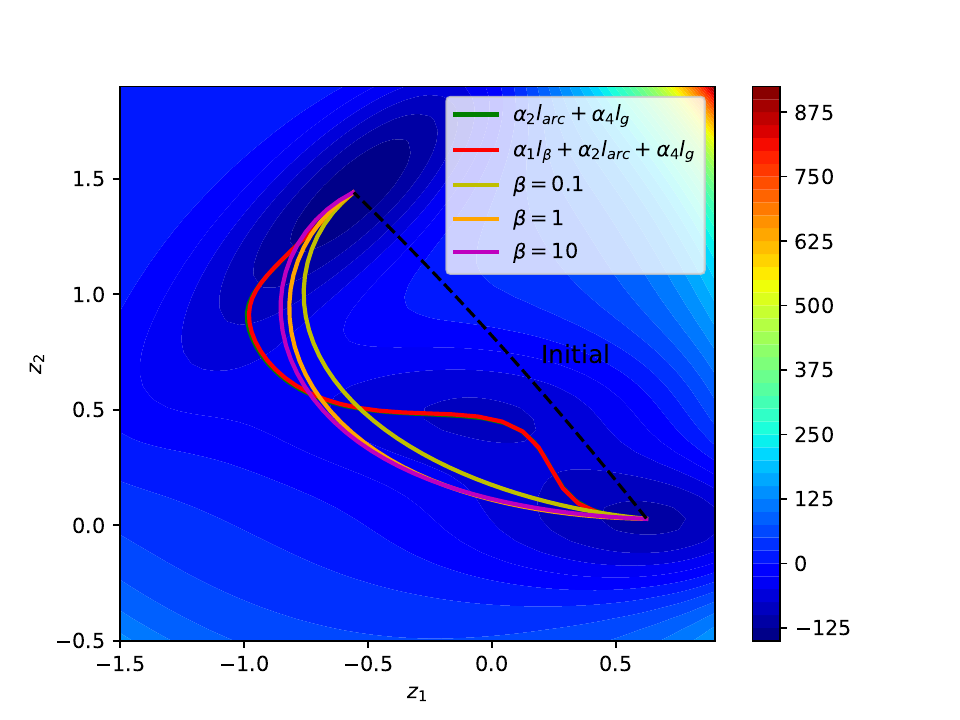}
    \end{minipage}
\caption{ The numerical MEP  as well as   MFP at three $\beta$ for the ten dimensional Muller potential problem. The plot is the projection of the path
$\varphi(x)$ onto  the $(z_1,z_2)$ plane. }
 \label{path_compare_mp10}
\end{figure}

\begin{figure*}[htbp]
\centering

          \includegraphics[width=.32\linewidth]{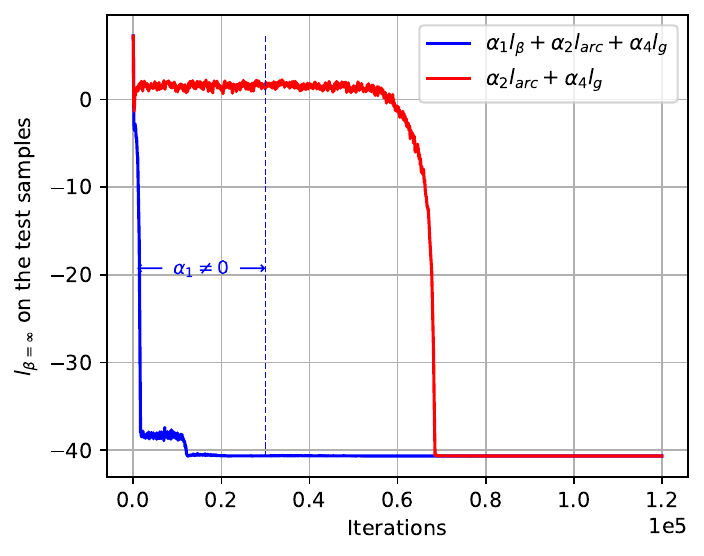}
  \includegraphics[width=.32\linewidth]{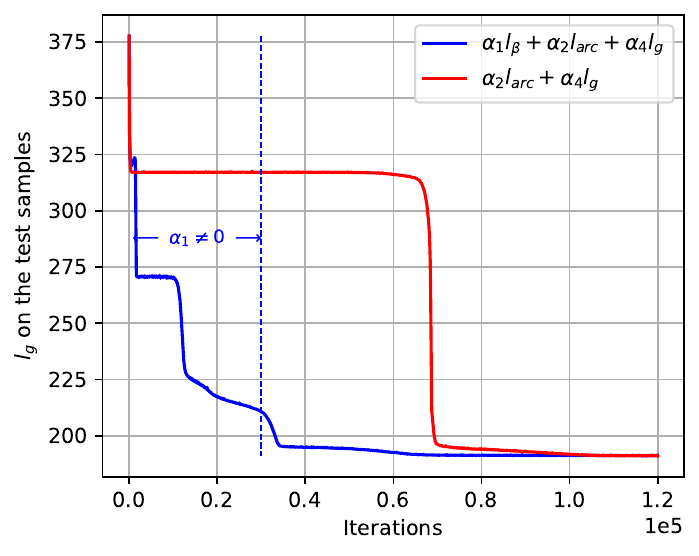}
 \includegraphics[width=.32\linewidth]{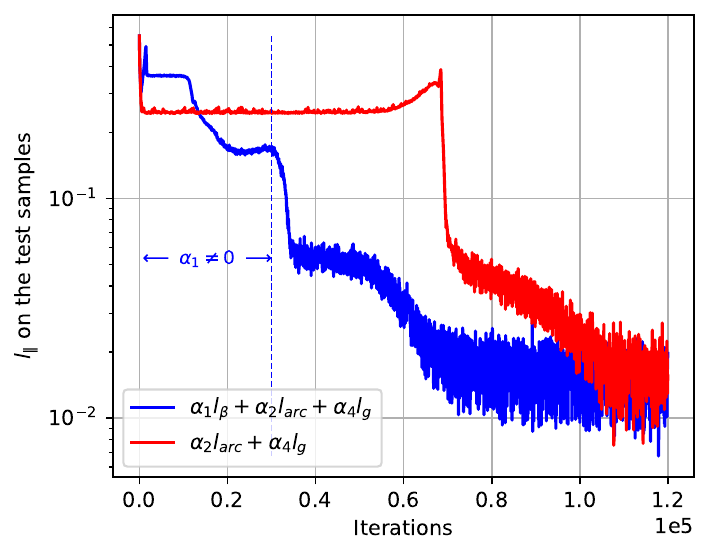}

\caption{The acceleration effect of the pre-training
by $\ell_\beta$ when using the geometric action $\ell_g$ to find the MEP for 10 dimensional Muller potential.
Three plots are the  decay of 
 three measurements of the path quality, measured by $\ell_{\beta=\infty}$ (i.e., $\max_s U(\varphi(s))$), $\ell_g$,  $\ell_\parallel$, respectively. 
The early stage of applying $\ell_\beta$ by setting a non-zero weight $\alpha_1$  help the training process from all
these three aspects. 
}
\label{fig:loss_compare_mp10}

\end{figure*}

The   Muller-Brown potential in two dimension   \cite{String2002} is
$$
    U_{M}(x_1,x_2) = \sum_{i=1}^4 D_i e^{a_i(x_1 - X_i)^2 + 
    b_i(x_1-X_i)(x_2-Y_i) + c_i(x_2-Y_i)^2},
$$ with the parameters 
$D_i, a_i, b_i, c_i$ and $X_i, Y_i$.
The   extension of this  potential in $\mathbb{R}^d$ is simply  set as  :\begin{equation}
\begin{split}
U_0(z_1,z_2,\cdots, z_d) :=U_{M}(z_1,z_2)+ 2\sum_{j=3}^d 
      z_j^2, 
\end{split}
\end{equation}
In our test, the potential  to compute the path  is 
 $U(x):= U_0(z)$ where $z=Qx$ where $Q\in \mathbb{R}^{d\times d}$ is a known orthonormal  matrix. 
$Q$ is chosen as the Q  matrix 
  by performing a QR factorization of a 
  random matrix with elements i.i.d. standard Gaussian random variables; this orthonormal matrix  $Q$ is  fixed later in the computation. $d=10$ is set in our test. 
  The potential has three local minima and two saddle points. We are interested in the transition path between the local minima.

The projection of the path $\varphi(x)$ is plotted in $(z_1,z_2)
$  and shown in  Figure \ref{path_compare_mp10} where we compare the $\beta$ max-flux paths and the minimum energy path. We observed that  the first saddle point with the high energy from the first local well on the top is important for  the max-flux path, while  the second intermediate state (the local well in the middle) is less related to the max-flux path.


We use this ten dimensional path to test 
the pre-training technique  by 
using the max-flux path loss to accelerate the action minimization. We include $\ell_\beta$ in the first 3000 iterations with the original action loss and then turn this loss off later.
We  monitor the accuracy of the path during the training steps  in Figure \ref{fig:loss_compare_mp10} by plotting the performance indicators of the maximum potential along the path 
($\ell_{\beta=\infty}$), the geometric action value $\ell_g$ and the tangent condition $\ell_\parallel$, respectively, in three subplots.  
For all three performance indicators,   we see the effectiveness of  the pre-training in accelerating the training efficiency.


\subsection{Comparison of Network Structures}

\begin{figure}[htbp]
\centering
\subfigure[The decay of three losses in terms of the number of 
iterations in training. ]
{
 	\begin{minipage}[b]{0.985\linewidth}
        \centering
        \label{MP_pth1}
        \includegraphics[width=.32\linewidth]{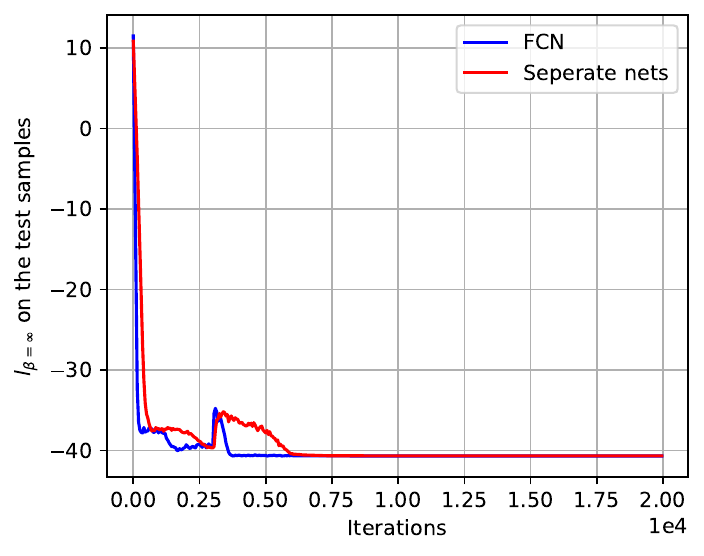}
           \includegraphics[width=.32\linewidth]{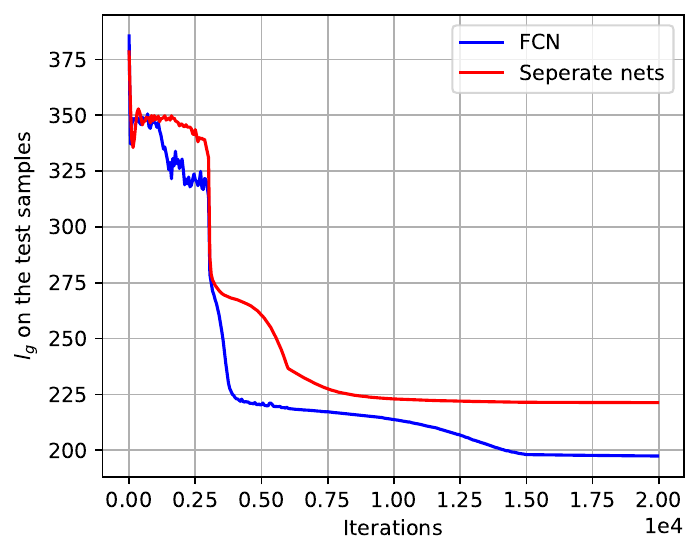}
           \includegraphics[width=.335\linewidth,height=0.255\linewidth]{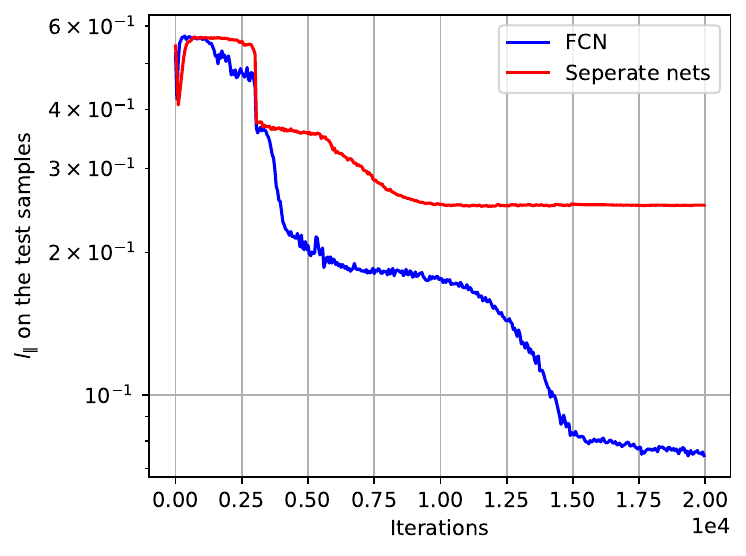}
    \end{minipage}
}
 
\subfigure[The   decay of three losses in terms of the CPU times]
{
 	\begin{minipage}[b]{ 0.985\linewidth}
        \centering
        \label{MP_pth2}
        \includegraphics[width=.32\linewidth]{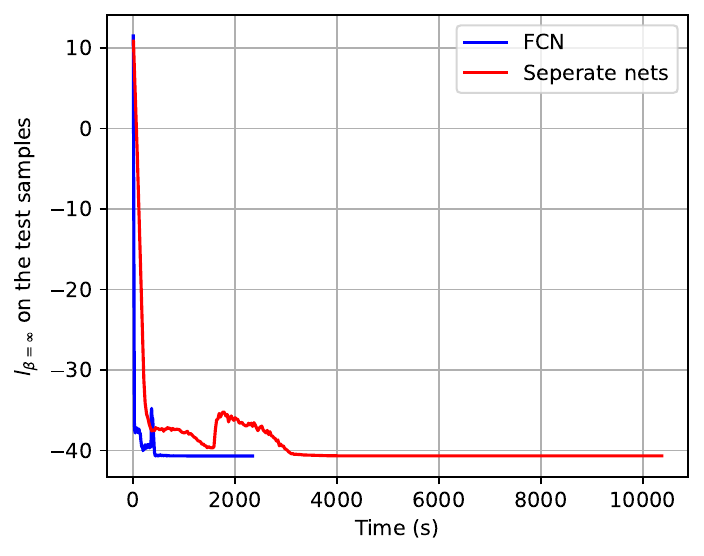}
             \includegraphics[width=.32\linewidth]{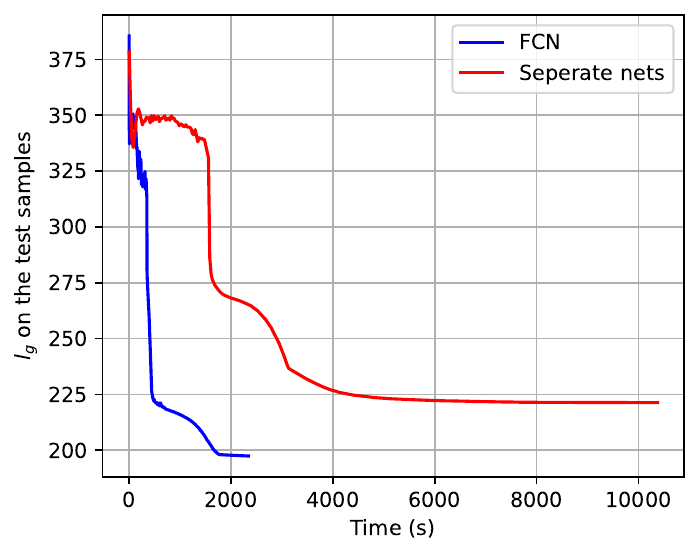}
         \includegraphics[width=.335\linewidth,height=0.255\linewidth]{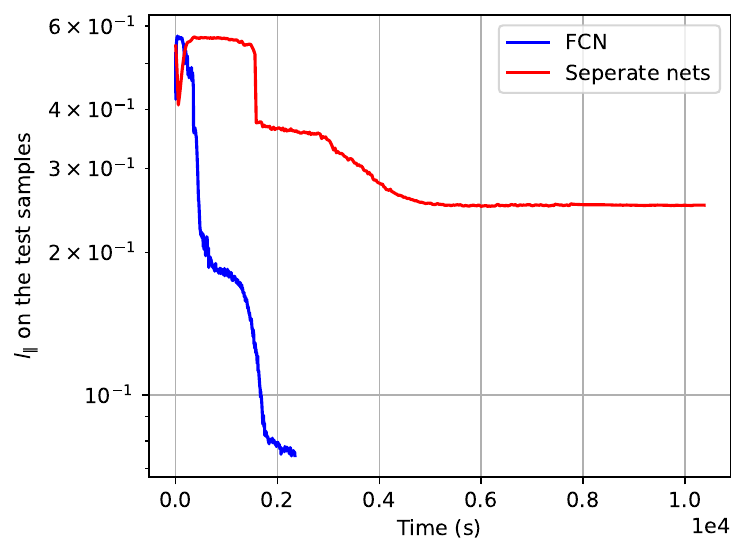}

    \end{minipage}
}
 \caption{Comparison of fully-connected neural network and separate network structures  in  Figure \ref{fig:NN}, tested on the  10 dimensional Muller potential problem. 
 Both networks have  the same number of parameters,  Three measurements
 of the path quality are the same as  in  Figure \ref{fig:loss_compare_mp10}.
 The plots are for the iteration number and 
the actual CPU running time up to 20000 total iterations, respectively.
All training settings are the same for the two different networks. The CPU times here include the calculation of test loss values in the plot. The computation is on Tesla T4 GPU  at Google Colaboratory ).
}\label{fig:nets_loss}
\end{figure}


In our previous numerical tests,  the path $\varphi: [0,1]\to \Real^d$  was represented by  the fully connected neural network   as shown in Figure \ref{fig:fcn}.  This architecture  
can be naturally extended to the function space, where $\varphi(\xb,s)$ is a function from $\Real^{n+1}\to \Real$. 
Before  we move to this setting of Hilbert space in the next section,  
we will make the comparison of the two   network structures  mentioned in Section \ref{ss:NN}
and Figure   \ref{fig:NN} for the finite dimension setting first  by testing on the ten-dimensional Muller potential problem mentioned above in Section \ref{ss:MP}.

In order to show  that the single network   architecture  in Figure \ref{fig:fcn} is not only more efficient but also provides more accurate results, we compare it with a network with $d$ separate sub-networks (as shown in  \ref{fig:separate}). Each sub-network  approximates one dimension of $\varphi$. 
For a fair comparison, we set the same
total number of neural network parameters  for two networks.  Specifically, the number of neurons at each layer of the fully-connected network (FCN) is $[1,20,20,20,20,10]$, while  the number of neurons for each layer of the sub-networks
for the separate network  is $[1,6,6,6,6,1]$. This suggests that the FCN has $20\times 20 +20=420$ parameters for each hidden layer and $1\times 20+20+4\times 420 +20\times 10=1920$ parameters totally.
The separate network with 10   sub-nets has $(6\times6+6)\times 10=420$ parameters  for one hidden layer and $(1\times6+6)\times 10+4\times 420+(1\times6+6)\times 10=1920$ parameters. 
 These two networks are then optimized
 for the loss function $  \alpha_1 \ell_\beta+\alpha_2   \ell_{arc} + \alpha_4 \ell_g$, where
  $\alpha_1 =1, \alpha_2=0.1, \alpha_4=0.001$ for the first $5000$ iteration and $\alpha_1 =0.1, \alpha_2=0.1, \alpha_g=10$ for the rest iterations. Both networks are trained for 20000 iterations.
  All  trainings  settings are the same for both network structures. The learning rate is  $1\times 10^{-4}$.
  
  \begin{figure}[htbp]
\centering
 	\begin{minipage}[b]{\linewidth}
        \centering
    \includegraphics[width=.8\linewidth]{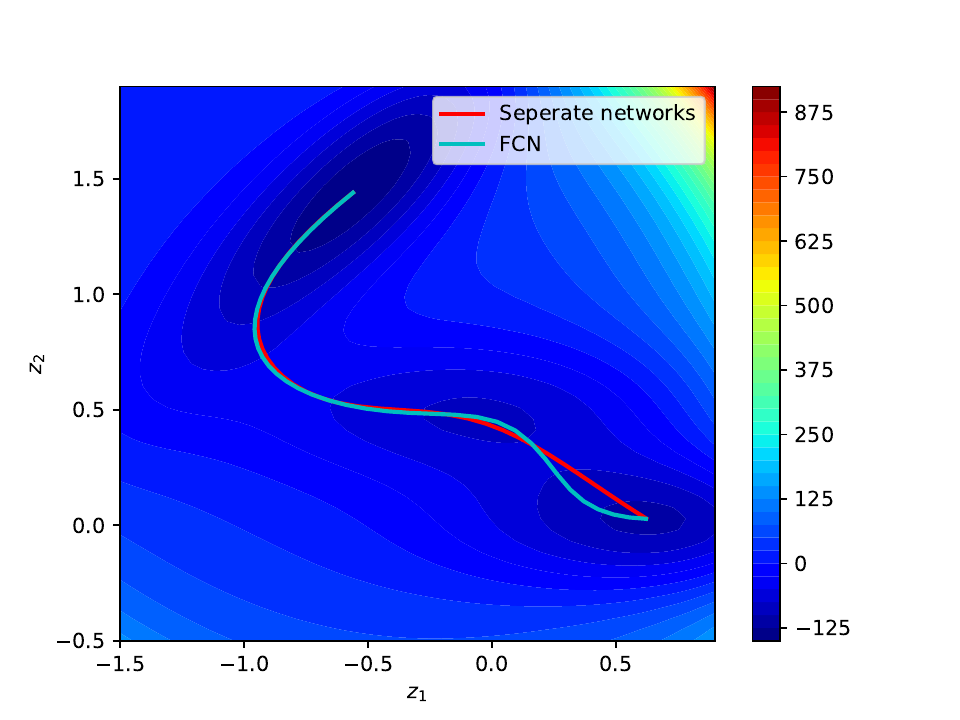}
    \end{minipage}
\caption{ The comparison of  the numerical  MEP  from the FCN and the separate $10$ sub-networks.}
 \label{fig:nets_path}
 \end{figure}

We record the performance measured by three losses  at each iteration and  the   CPU time,   shown in Figure \ref{fig:nets_loss}.  
It is clear that the use of one neural network  
is highly recommended due to its better optimization efficiency and lower computational cost.
Figure \ref{fig:nets_path} also shows the comparison of the final results of the MEP  as well. It can be concluded that the use of a single neural network by sharing the parameters across the components of the path is more efficient  and more accurate.

\section{$n$-dimensional  Ginzburg-Landau functional}

We now consider  the paths for the Ginzburg-Landau functional defined 
by \eqref{GL-ef}     on the unit hyper-cube $\Omega=[0,1]^n$ in
any $n$ dimensional space.
As discussed in Section \ref{ssec:PDE}, the path is now represented 
by $\varphi(s,\xb)$. The two- and four- dimensional problems  ($n=2,4$) are computed
and demonstrated below with two values of $\kappa$. 

\subsection{$n=2$ }

\begin{figure*}[htbp]     \includegraphics[width=\linewidth]{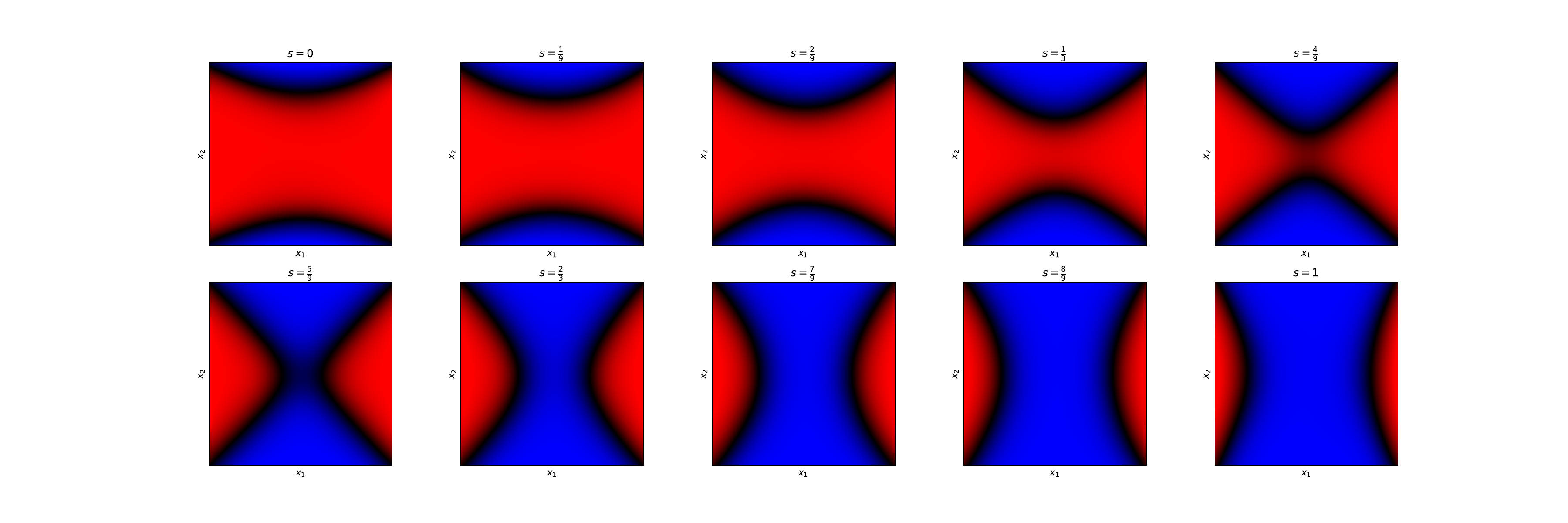}
    \caption{The numerical MEP of 2-dim  Ginzburg-Landau functional  ($\kappa=0.08$).}\label{fig:d2_s}
 \end{figure*}
\begin{figure*}[htbp]
     \includegraphics[width=\linewidth]{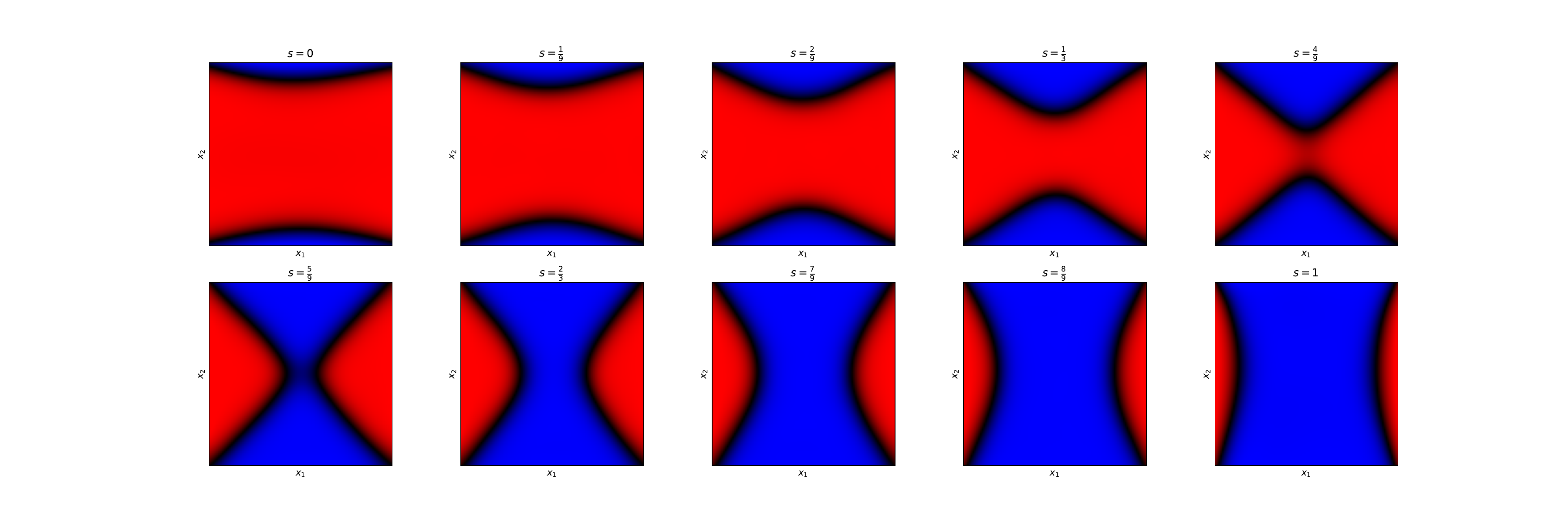}
    \caption{The numerical  MEP of 2-dim Ginzburg-Landau functional ($\kappa=0.05$) 
    obtained by using the  $\ell_\beta$ pre-training.}\label{fig:d2_s_beta}
\end{figure*}
\begin{figure*}[htbp]
     \includegraphics[width=\linewidth]{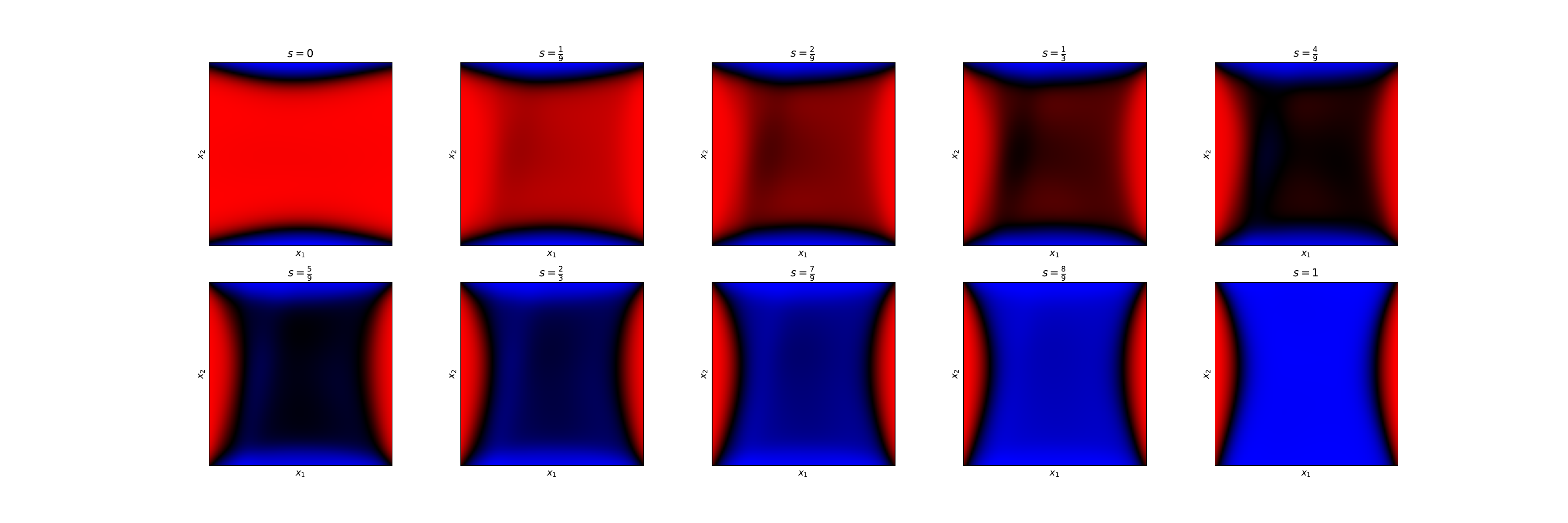}
    \caption{The (bad) numerical  MEP of 2-dim Ginzburg-Landau functional ($\kappa=0.05$), without the pre-training of  max-flux path using $\ell_\beta$.}\label{fig:d2_s0}
\end{figure*}

We take the following Dirichlet  boundary condition on $(0,1)^2$ for the function $u$:
$u(x_1=0,x_2)=u(x_1=1,x_2)=1$ and $u(x_1,x_2=0)=u(x_1,x_2=1)=-1$.
The two local minimizers $u_a$ and $u_b$ of $E(u)$ are computed  beforehand with the neural networks of width $[2,32,32,32,1]$.  There is a symmetry such that  $u_a(x_1,x_2) =-u_b(x_2,x_1)$. Our neural networks are the five-layer feedforward neural networks with  the widths $[3,100,100,100,1]$.
The activation function is $\tanh$.

We calculated the pathways from $u_a$ to $u_b$. 
 The path $\varphi(s, \xb)$ needs to satisfy the Dirichlet boundary condition for all $s$, 
which is  enforced in the penalty form on the top of the existing loss function. It
 satisfies   $\varphi(s=0,\xb)=u_a(\xb)$ and $\varphi(s=1,\xb)=u_b(\xb)$
at the two endpoints,  enforced by the neural network structure \eqref{NN} and \eqref{line-init}.
The training samples now are both in $\xb$ and in $s$. 
In the space domain $(0,1)^2$, $50\times 50$ points are sampled uniformly.
$s$ is uniformly sampled with $50$ points.  These settings are the same for 
the two values of 
$\kappa=0.08$ and $\kappa=0.05$ tested here.
We compute the MEPs  by the minimum action method by minimizing 
$\ell_g$. The smaller the $\kappa$, the more challenging it is for the optimization to find the path.
 As we have  witnessed in previous examples,  
  the pre-training with $\ell_\beta$ can significantly help improve training efficiency.

\begin{figure}[htbp]
\centering
\subfigure[$\kappa=0.08$]{
    \includegraphics[width=0.46\linewidth]{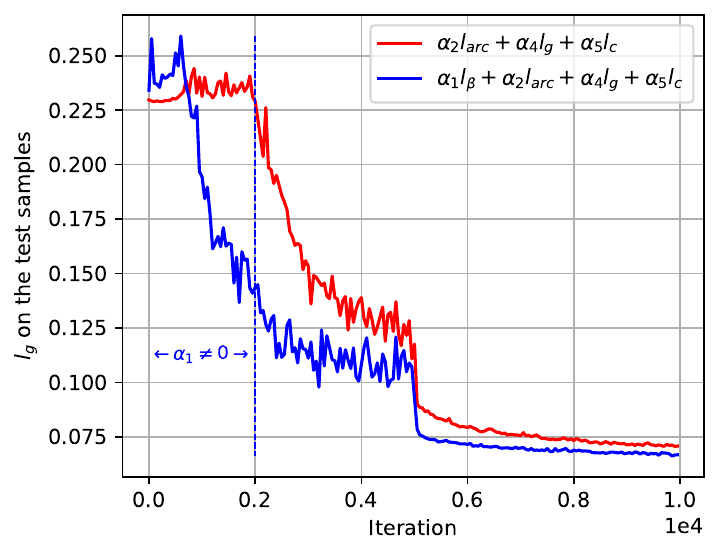}}
    \subfigure[$\kappa=0.05$]{
        \includegraphics[width=0.46\linewidth]{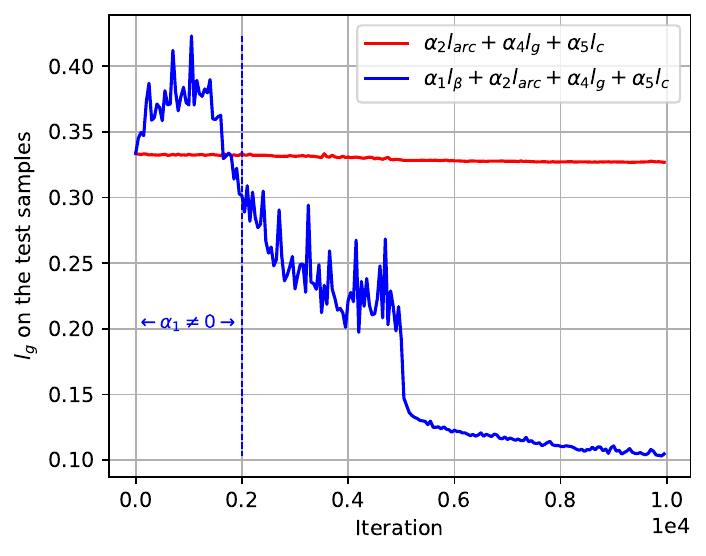}}
    \caption{ The comparison for the geometric action loss during the training for   computing the MEP   of 2-dim Ginzburg-Landau functional, with (blue) and without (red) pre-training by max-flux loss $\ell_\beta$, respectively. $\ell_c$ means the penalty loss for the Dirichlet boundary condition.}\label{fig:d2_k008_com}
\end{figure}

We first show the  MEPs at these two 
$\kappa=0.08$ and $\kappa=0.05$ in Figure \ref{fig:d2_s} and Figure \ref{fig:d2_s_beta}, respectively.  The red, black and blue denote  the value $+1$, $0$ and  $-1$ respectively. 
Note that  Figure  \ref{fig:d2_s_beta} at $\kappa=0.05$ is obtained with the pre-training  by adding $\alpha_1\ell_\beta$ in the first stage ($\alpha_1=
10$ and $\beta=10$).
As an ablation study, Figure \ref{fig:d2_s0} used only the loss of $\ell_g$ without the pre-training techniques.  The result in Figure \ref{fig:d2_s0} is much worse than that in Figure  \ref{fig:d2_s_beta}.  To validate how the pre-training
using the max-flux loss accelerates the action minimization,  we also plot in Figure 
\ref{fig:d2_k008_com} the  value of the geometric action 
$\ell_g$ as  the training  progresses, with or without the pre-training
($\alpha_1$).
The pre-training is implemented by using $\alpha_1$ for the $\ell_\beta$ loss 
for the first two thousand  iterations. After that, this max-flux loss $\ell_\beta$ is then removed to compute the MEP precisely.  In both cases, we see the effectiveness of adding $\ell_\beta$ in the pre-training. Particularly for the smaller   $\kappa=0.05$,  the direct minimization which only relies on the action loss is more challenging than the case at $\kappa=0.08$, while the pre-training does help find the right solution.

\subsection{$n=4$}

\begin{figure*}[htbp]
\centering
\subfigure[s=0]
{
 	\begin{minipage}[b]{0.3\linewidth}
        \centering
\includegraphics[width=1\linewidth]{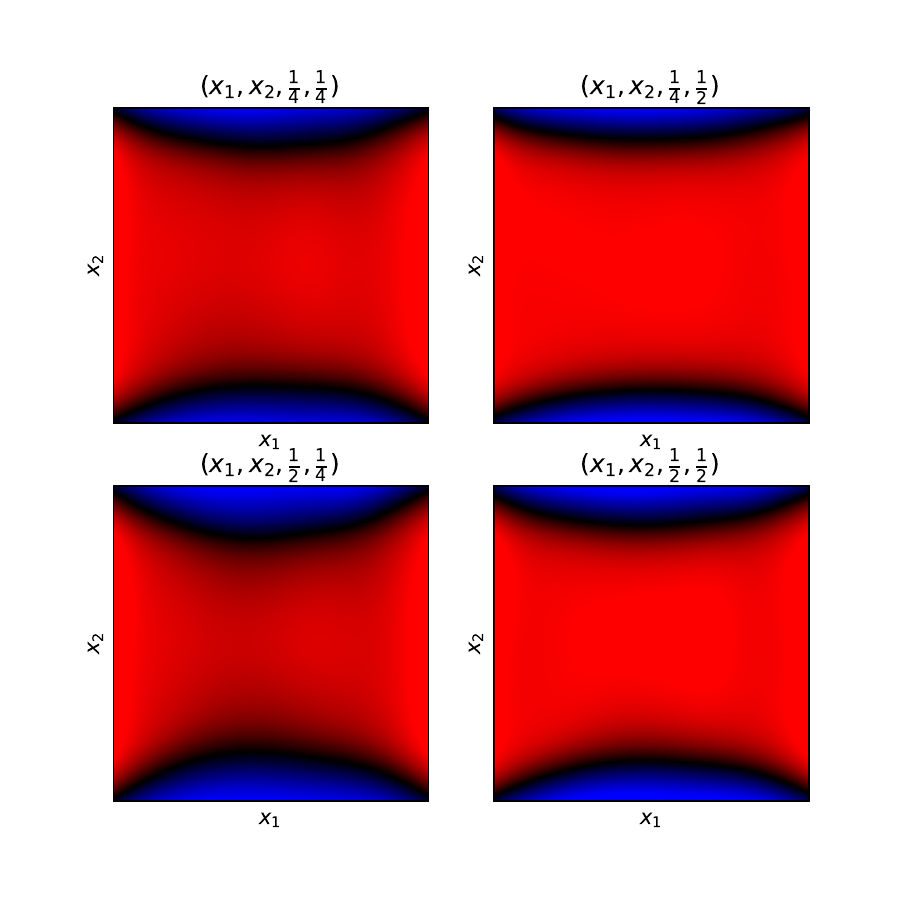}
    \end{minipage}
}
\subfigure[s=0.25]
{
 	\begin{minipage}[b]{.3\linewidth}
        \centering
        \includegraphics[width=1\linewidth]{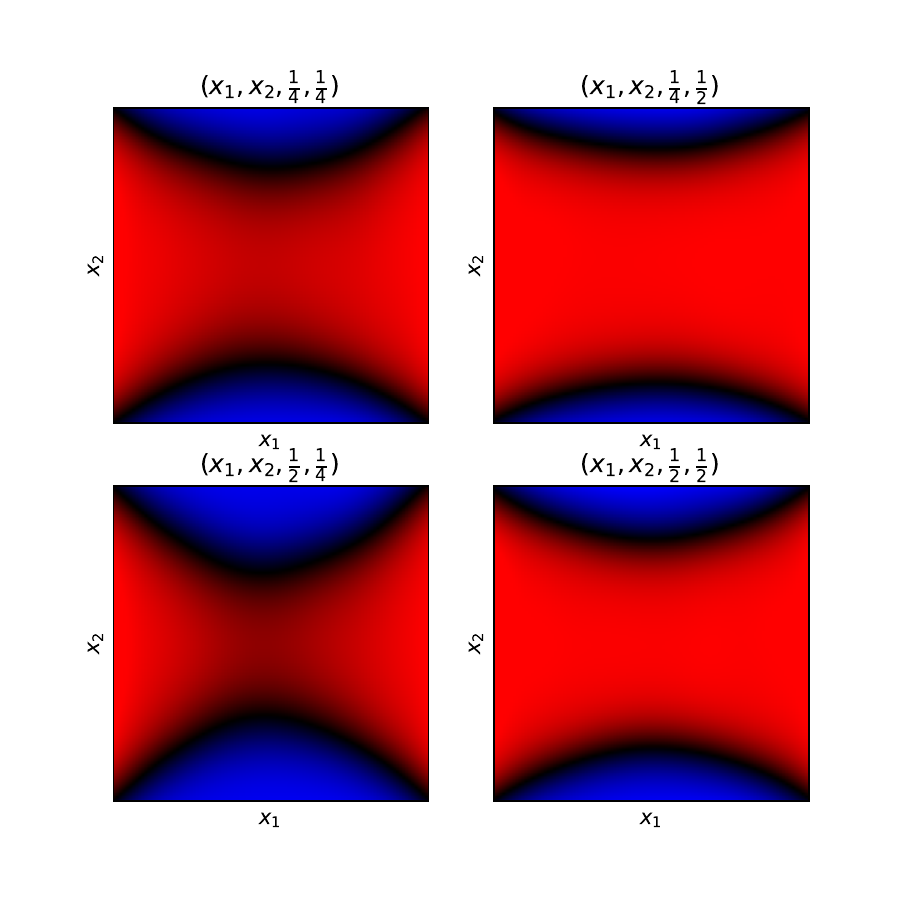}
    \end{minipage} 
}
\subfigure[s=0.5]
{
 	\begin{minipage}[b]{0.3\linewidth}
        \centering
\includegraphics[width=1\linewidth]{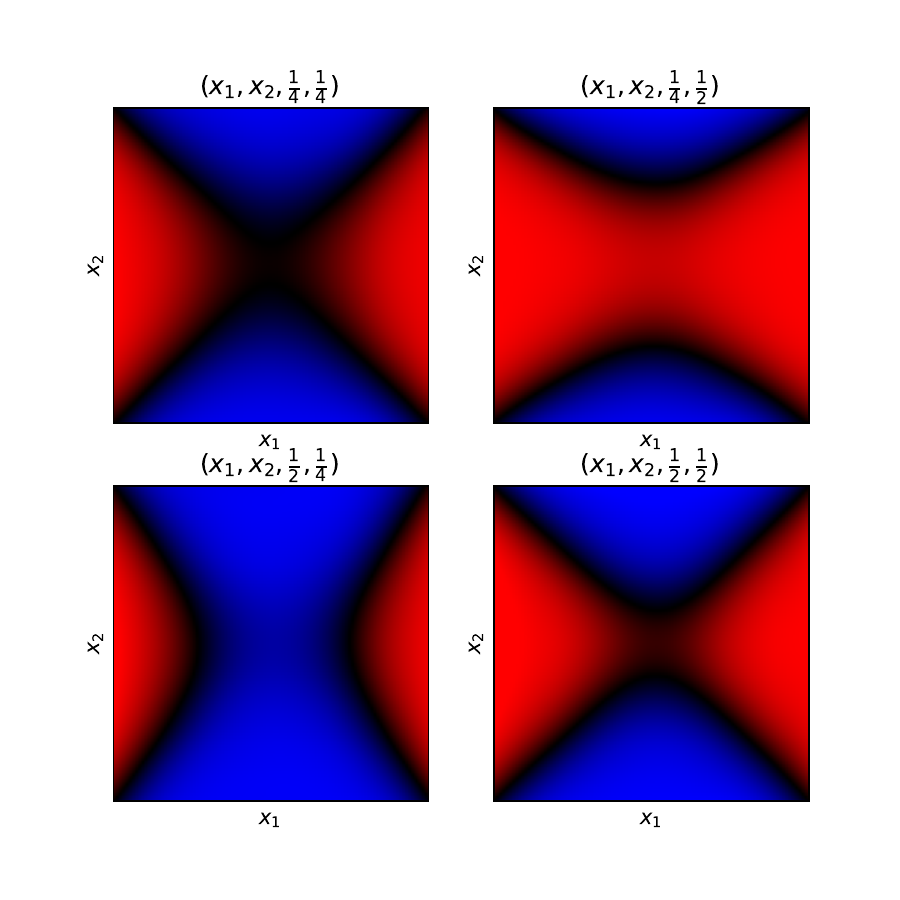}
    \end{minipage}
}

\subfigure[s=0.75]
{
 	\begin{minipage}[b]{.3\linewidth}
        \centering
        \includegraphics[width=1\linewidth]{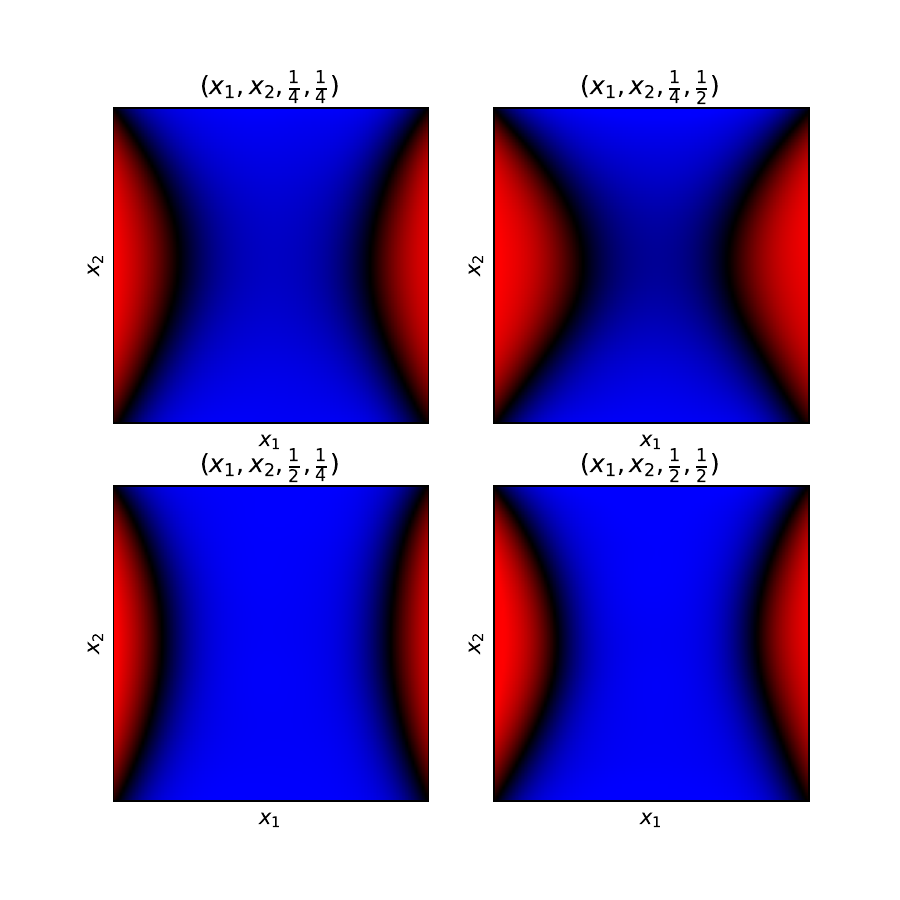}
    \end{minipage} 
}
\subfigure[s=1]
{
 	\begin{minipage}[b]{.3\linewidth}
        \centering
        \includegraphics[width=1\linewidth]{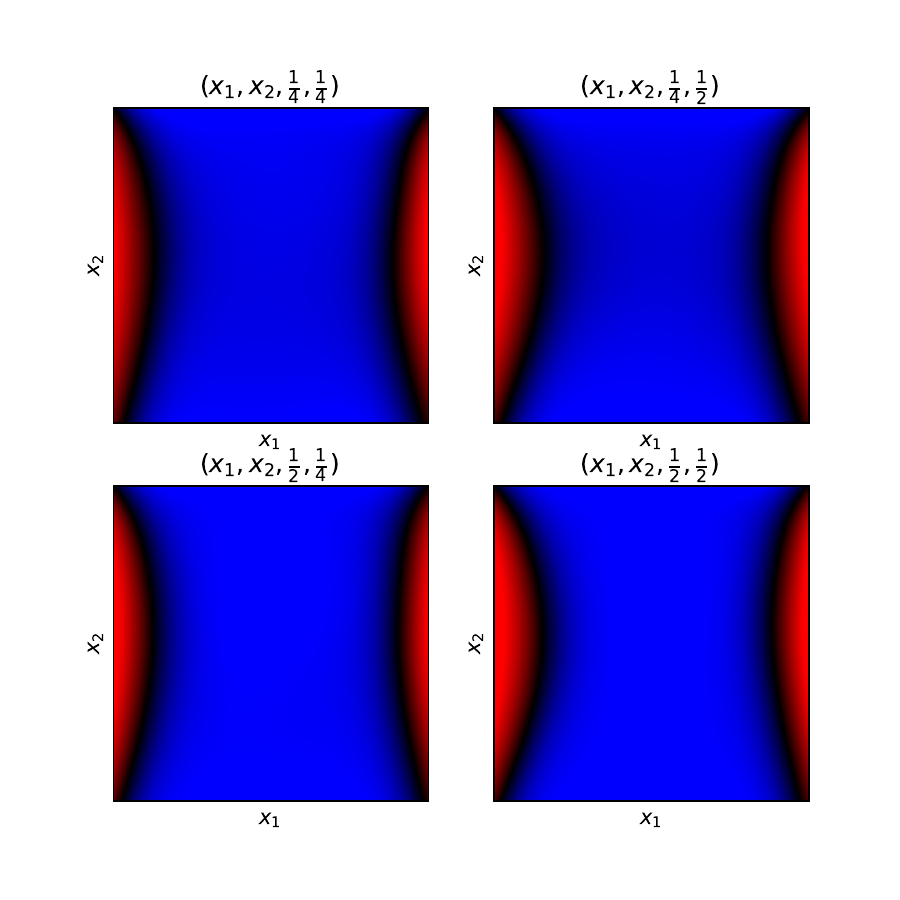}
    \end{minipage} 
}
\subfigure[$\ell_g$ loss]
{
 	\begin{minipage}[b]{.3\linewidth}
        \centering
    \includegraphics[width=1\linewidth,height=.86\textwidth ]{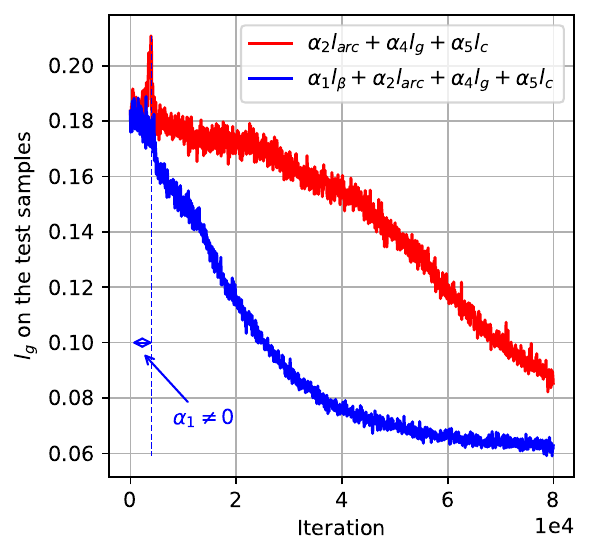}
    \end{minipage} 
}

\caption{Two dimensional visualization of the  numerical MEP ((a)-(e)), plotted on some cross-sections, of four dimensional Ginzburg-Landau functional  with the pre-training technique ($\kappa=0.08$), (f): The comparison for the geometric action $\ell_g$ during the training, with (blue) and without (red) pre-training by max-flux loss, respectively.
}\label{fig:loss_compare_dim4_k008}

\end{figure*}


\begin{figure*}[htbp]
\centering
\subfigure[s=0]
{
 	\begin{minipage}[b]{0.3\linewidth}
        \centering
\includegraphics[width=1\linewidth]{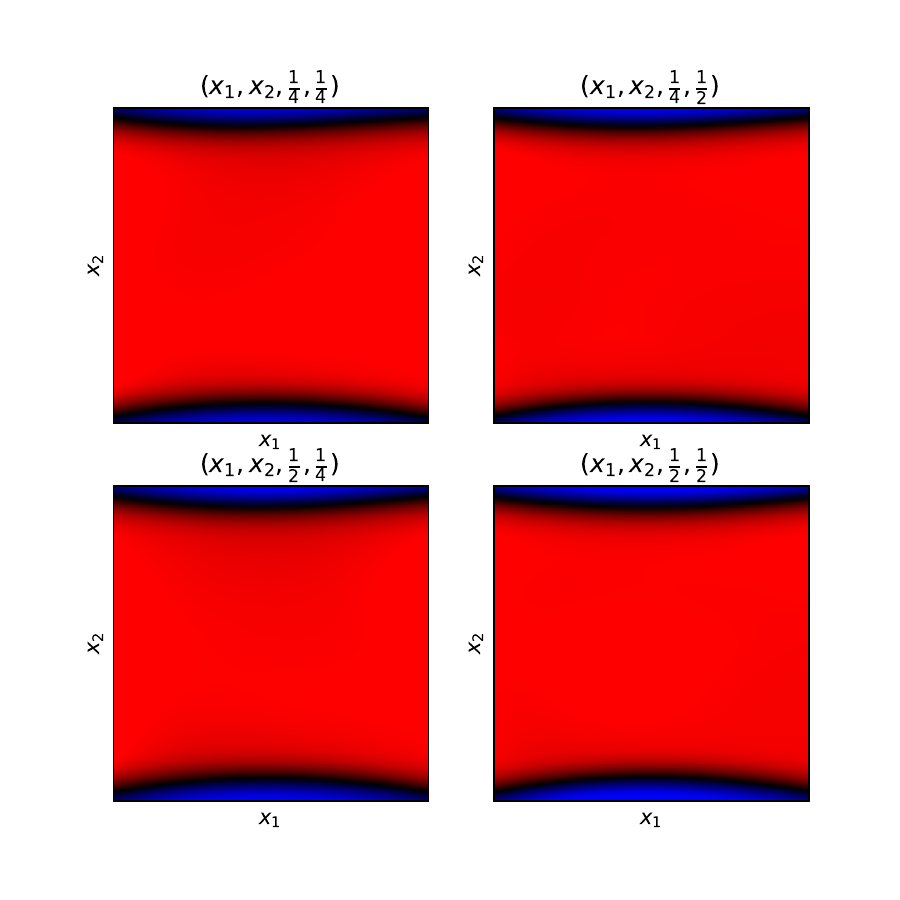}
    \end{minipage}
}
\subfigure[s=0.25]
{
 	\begin{minipage}[b]{.3\linewidth}
        \centering
        \includegraphics[width=1\linewidth]{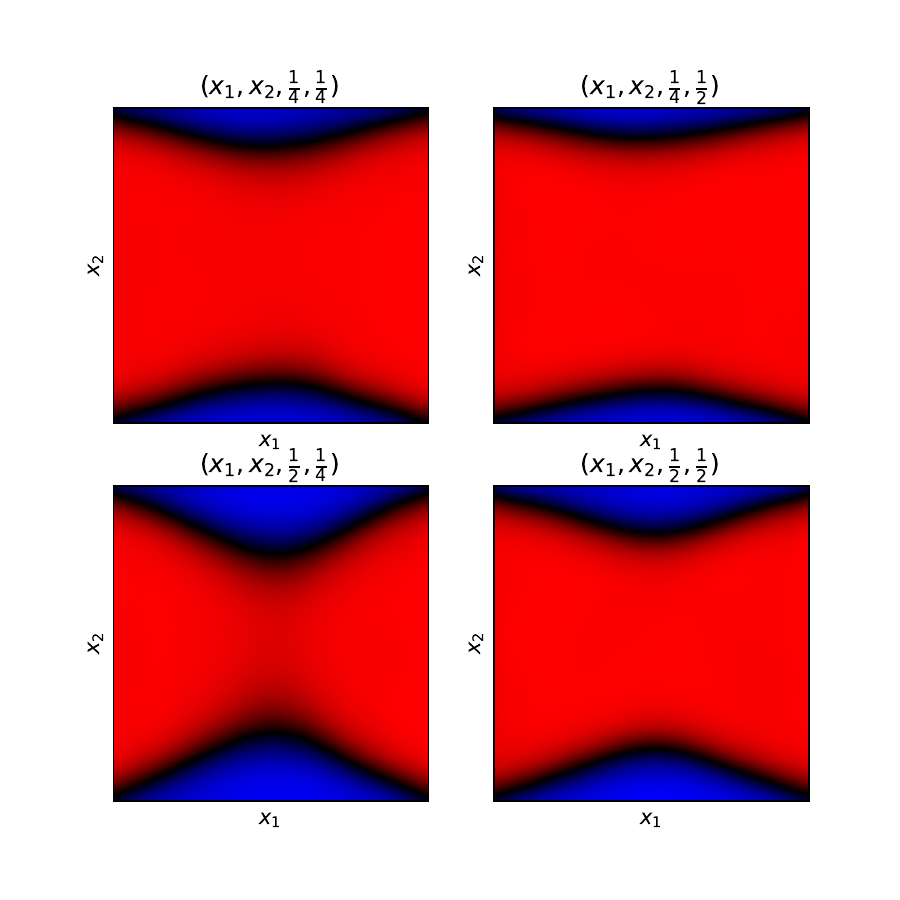}
    \end{minipage} 
}
\subfigure[s=0.5]
{
 	\begin{minipage}[b]{0.3\linewidth}
        \centering
\includegraphics[width=1\linewidth]{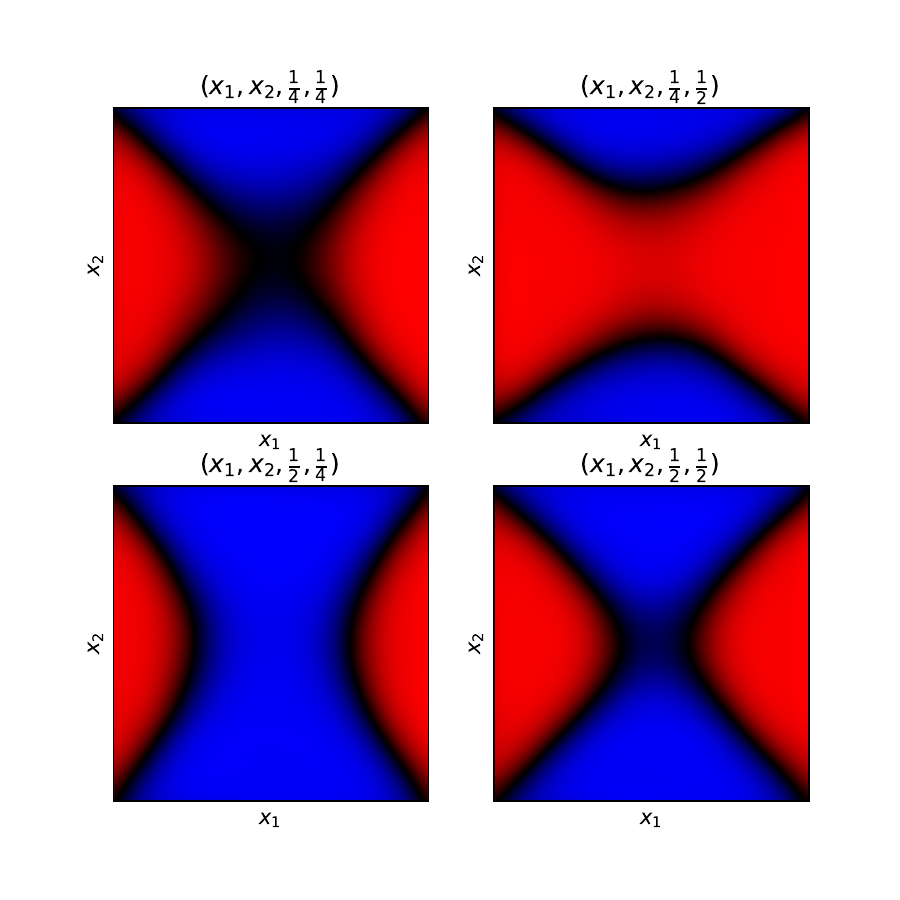}
    \end{minipage}
}
\subfigure[s=0.75]
{
 	\begin{minipage}[b]{.3\linewidth}
        \centering
        \includegraphics[width=1\linewidth]{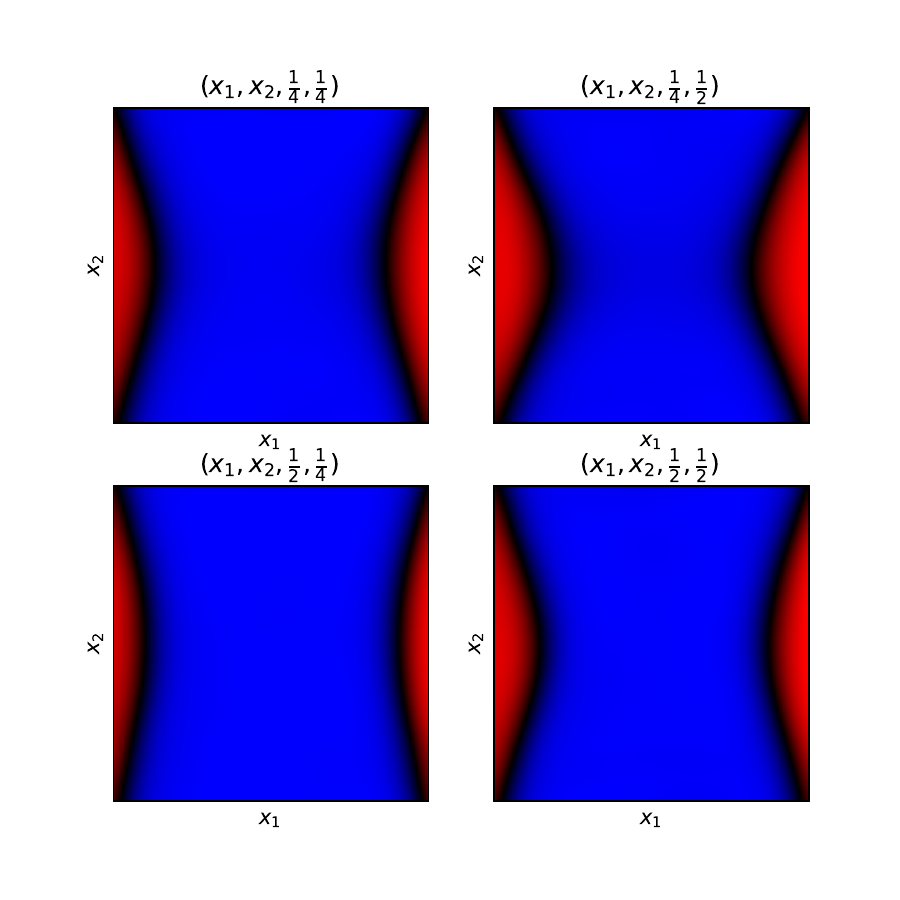}
    \end{minipage} 
}
\subfigure[s=1]
{
 	\begin{minipage}[b]{.3\linewidth}
        \centering
        \includegraphics[width=1\linewidth]{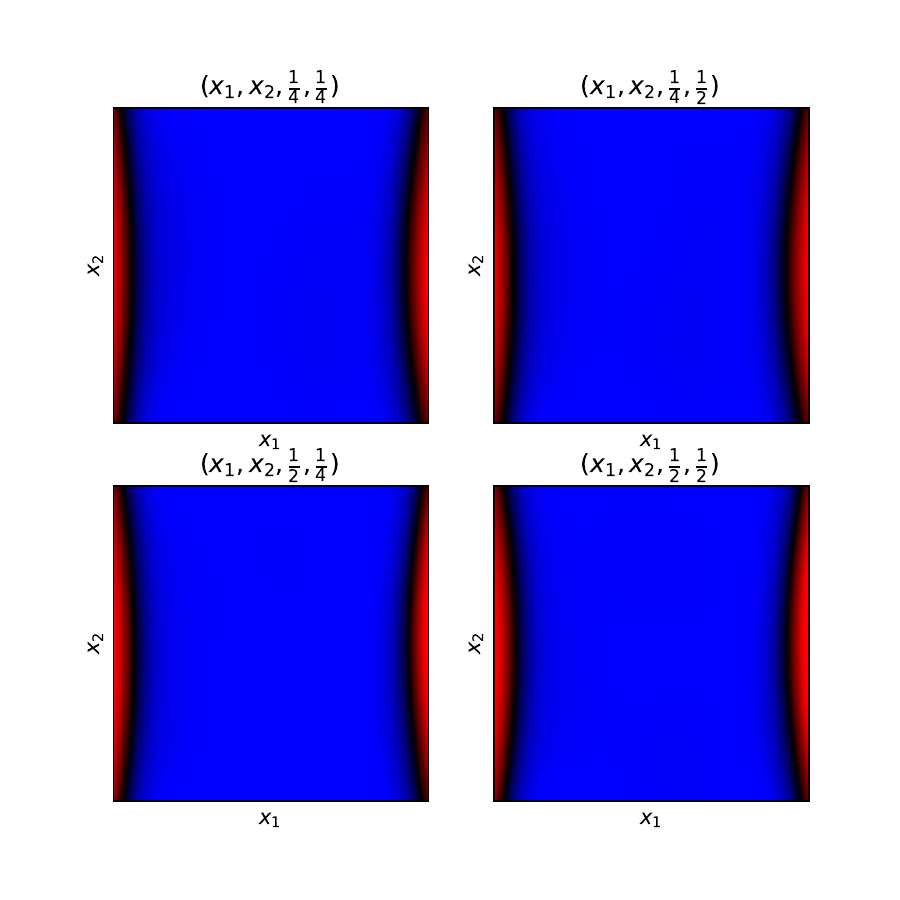}
    \end{minipage} 
}
\subfigure[ $\ell_g$ loss]
{
 	\begin{minipage}[b]{.3\linewidth}
        \centering
    \includegraphics[width= 1\linewidth, height=.86\textwidth]{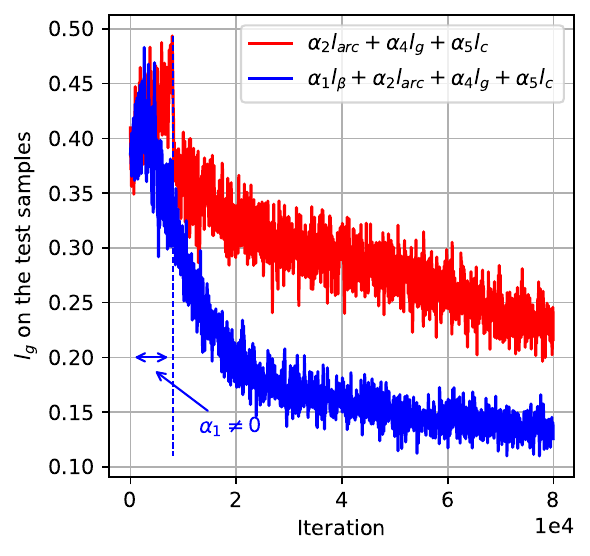}
    \end{minipage} 
}
\caption{Two dimensional visualization of the  numerical MEP ((a)-(e)), plotted on some cross-sections, of four dimensional  Ginzburg-Landau functional  with the pre-training technique ($\kappa=0.05$). (f): The comparison for the geometric action $\ell_g$ during the training, with (blue) and without (red) pre-training, respectively.)
}
\label{fig:loss_compare_dim4_k005_success}

\end{figure*}


\begin{figure*}[htbp!]
\centering
\subfigure[s=0.25]
{
 	\begin{minipage}[b]{.31\linewidth}
        \centering
        \includegraphics[width=1\linewidth]{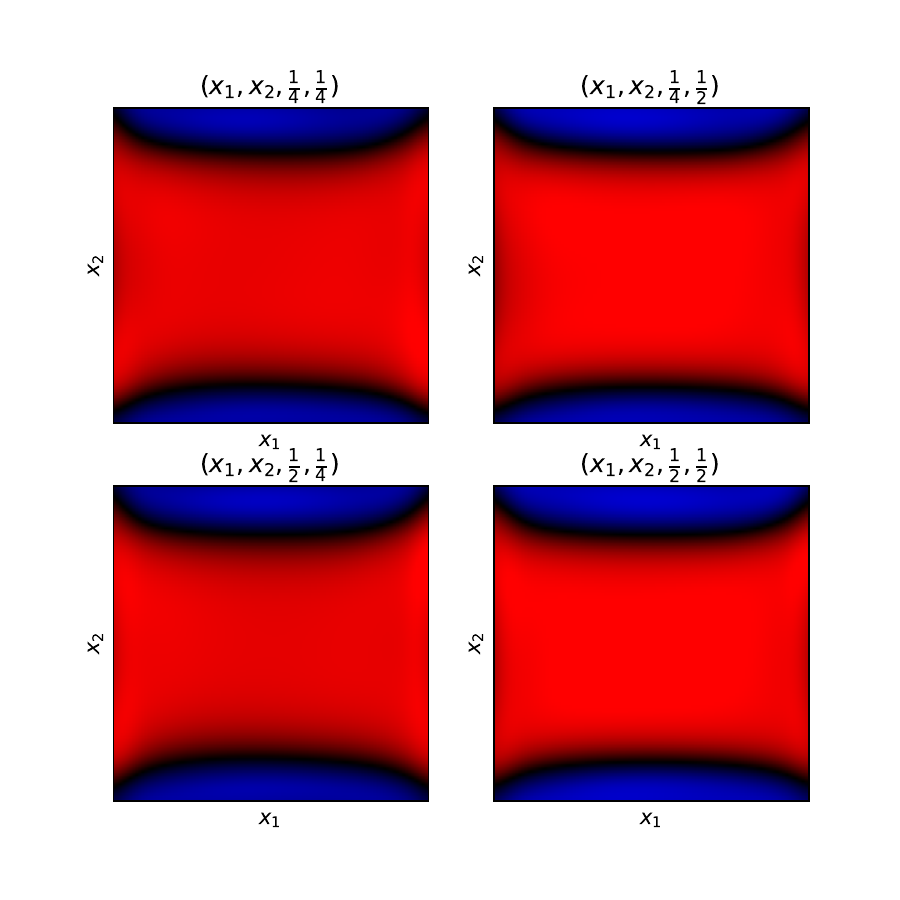}
    \end{minipage} 
}
\subfigure[s=0.5]
{
 	\begin{minipage}[b]{0.31\linewidth}
        \centering
\includegraphics[width=1\linewidth]{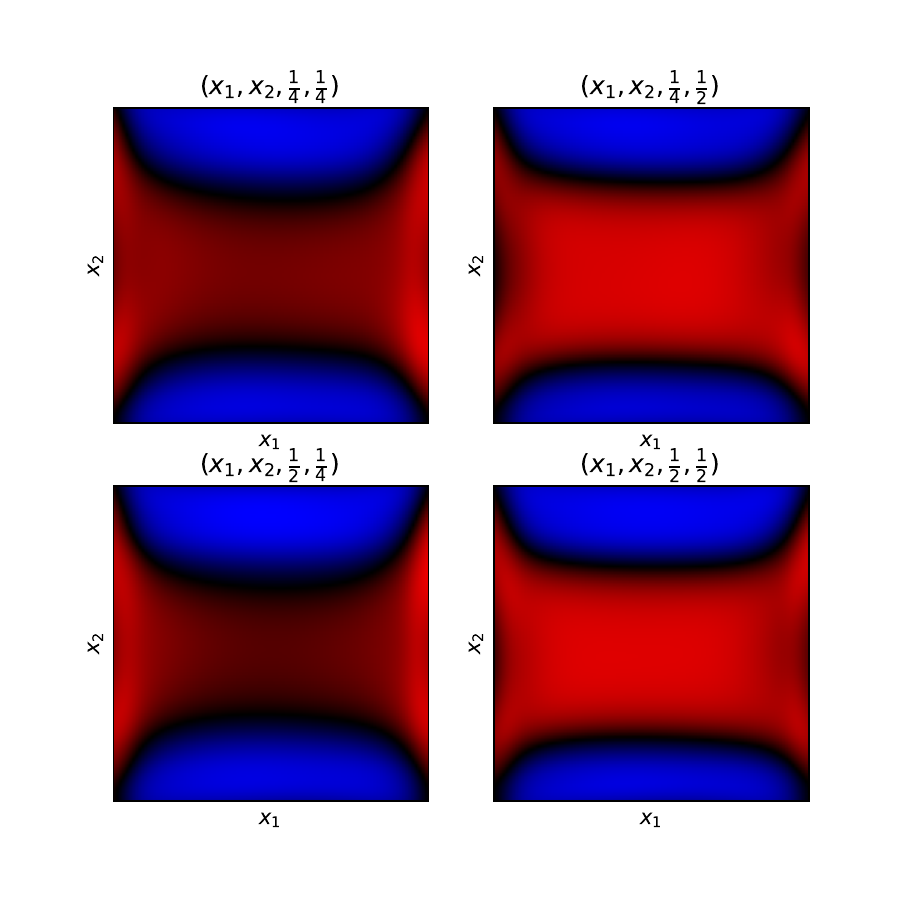}
    \end{minipage}
}
\subfigure[s=0.75]
{
 	\begin{minipage}[b]{.31\linewidth}
        \centering
        \includegraphics[width=1\linewidth]{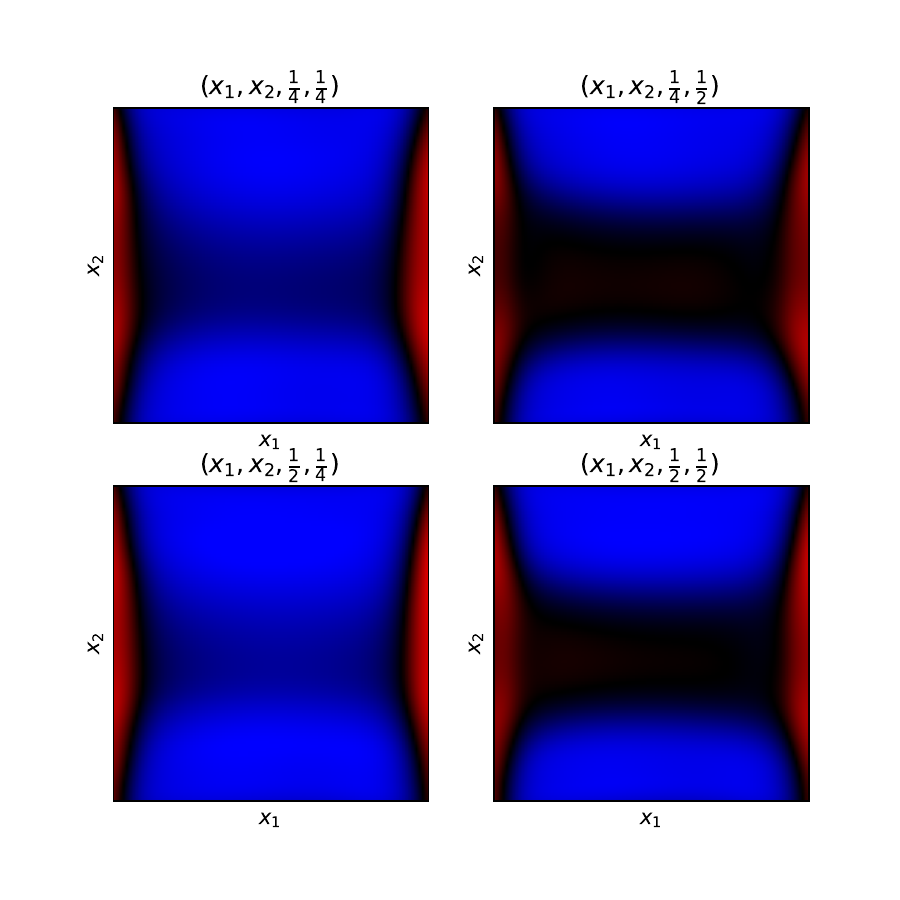}
    \end{minipage} 
}


\caption{The (bad) numerical  MEP of four dimensional Ginzburg-Landau functional ($\kappa=0.05$), without the pre-training   using $\ell_\beta$.
}
\label{fig:loss_compare_dim4_k005_fail}

\end{figure*}

To further show the feasibility and efficiency  of extending to high dimension,
we next test  the four-dimensional case $n=4$ for the   Ginzburg-Landau functional. 
The similar Dirichlet  boundary condition on the four dimensional hypercube $[0,1]^4$ 
is set: $u(\xb)=+1$ if $x_1(1-x_1)x_3(1-x_3)=0$,  and $u(\xb)=-1$ if $x_2(1-x_2)x_4(1-x_4)=0$.
The two local minimizer of $E$ is found 
by using the  fully-connected neural network  with the architecture 
$[4,64,64,64,64,1]$.
The StringNET is then a fully-connected network with an input of $5$ dimension and
an output of one dimension with four hidden layers:
$[5,100,100,100,100,1]$.

To visualize the results, we plot the four-dimensional profiles $\varphi(s,\xb)$ of path on selected $s$,   in terms of two variables  
$(x_1,x_2)$ at four cross-sections where 
$x_3$ and $x_4$ take values of  $1/4$ or $1/2$.
Figure \ref{fig:loss_compare_dim4_k008}
and Figure 
\ref{fig:loss_compare_dim4_k005_success}
show the results of the MEP at $\kappa=0.05$ and $\kappa=0.08$ respectively. The final panels in these two figures also confirm the improvement due to the pre-training technique. 
Figure \ref{fig:loss_compare_dim4_k005_fail}
shows the unsatisfactory result if the pre-training is not incorporated at $\kappa=0.05$.



\section{\label{sec:C}Conclusion}
In this paper, we build the StringNET, a method to calculate   transition pathways between two meta-stable states, including the minimum energy path and the maximum flux path.    StringNET leverages the deep learning techniques and is applicable to high-dimensional energy functionals.
It reformulates the gradient  flow dynamics in the string method into  variational problems for a continuous curve,   thereby avoiding the   reparametrization steps inherent in the string method.
We   present   strong numerical evidence against    using   the loss function  $\ell_\parallel$ based on the tangent condition, which has served as the   convergence condition in the original string method.
By exploiting the relationship between the $\beta$-dependent maximum flux path
and the minimum energy path,  we propose an acceleration  method 
that involves pre-training the maximum flux path in the early stage. We have demonstrated the 
feasibility and efficiency of this StringNET through several examples including the  four-dimensional Ginzburg-Landau equation.

Generally speaking, since   StringNET is based on   variational formulations,
it can be categorized within  the family of  minimum action methods. In fact, this method  for the MEP,  based on the loss $I_g$,  works in principle for any non-gradient dynamical systems  \cite{SIMONNET2023112349} by replacing 
$I_g$ with the true action \eqref{gaction}.  However,   pre-training 
using  the maximum flux loss $\ell_\beta$ involving the potential is not applicable to  such non-gradient systems.

\section*{Acknowledgments}
 GU acknowledges the support of NSFC 11901211 and the Natural Science Foundation of Top Talent of SZTU GDRC202137.    ZHOU acknowledges the support from 
 Hong Kong General Research Funds  (11308121, 11318522,   11308323),  and  the NSFC/RGC Joint Research Scheme [RGC Project No. N-CityU102/20 and NSFC Project No.
12061160462

\appendix


\section{First Variation 
in Curve space}
Let a smooth curve $C$ be represented by a function
$\mathbf{r}
(t):[0,1]\to \Real^d$, and define
the functional $E$ as the line integral of a function $F$  on this curve $C$: 
$$ E[\mathbf{r}] = \int_C F(\mathbf{r})\d\mathtt{s} = \int_0^1 F(\mathbf{r}(t)) ~|\mathbf{r}'(t)|\d t$$
where $F:\Real^d\to \Real$ is lower semicontinuous,
$\d \mathtt{s}$ is the arc length parameter and $t$ is an arbitrary parameter for the curve. $E$ is defined geometrically since the different parametrization of the curve will give the same value of $E$. 

Consider the infinitesimal perturbation of the curve $\mathbf{r}(t) \to \mathbf{r}(t) +\epsilon \delta \mathbf{r}(t)$, $t\in[0,1]$, where $\delta \mathbf{r}(0)=\delta \mathbf{r}(1)=0$  since the endpoints are fixed.
Then the first variation of $E$ is computed:
\begin{align*}
    & \frac{\d }{\d \epsilon}E(\mathbf{r}+ \epsilon \delta \mathbf{r}) \\
    = &\int_0^1 \nabla F(\mathbf{r})\cdot \delta \mathbf{r} ~\abs{\mathbf{r}'} 
     \d t
    +\int_0^1  F(\mathbf{r}(t))\frac{\mathbf{r}'(t)}{|\mathbf{r}'(t)|} \cdot \delta r'(t)\d t
    \\
     = & \int_0^1 \nabla F(\mathbf{r})\cdot \delta \mathbf{r} ~\abs{\mathbf{r}'} 
     \d t
   -\int_0^1  \frac{\d}{\d t} \qty( \frac{F(\mathbf{r}(t))\mathbf{r}'(t)}{|\mathbf{r}'(t)|} ) \cdot \delta \mathbf{r}(t)\d t
   \\
   =  & \int_0^1 \nabla F(\mathbf{r})\cdot \delta \mathbf{r} ~\abs{\mathbf{r}'} 
     \d t
   -\int_0^1  \frac{\d}{\d t} \qty(  F(\mathbf{r}(t))\boldsymbol{\tau}(t)  ) \cdot \delta \mathbf{r}(t)\d t
   \\
   =  & \int_0^1 \nabla F(\mathbf{r})\cdot \delta \mathbf{r} ~\abs{\mathbf{r}'} 
     \d t
   -\int_0^1    \qty(  \nabla F(\mathbf{r})
   \cdot \mathbf{r}' ) \qty[ \boldsymbol{\tau}\cdot \delta \mathbf{r}]\d t
   \\ \qquad &
   -\int     F(\mathbf{r} )~ \boldsymbol{\tau}'   \cdot \delta \mathbf{r}\d t,
\end{align*}
where $\mathbf{r}'(t)$ is the derivative w.r.t $t$, $\nabla F$ is the gradient of $F$
and $\boldsymbol{\tau}$ is the unit tangent vector.


    Now we obtain   the Euler-Lagrangian equation for the minimizer of $E[\mathbf{r}]$:
    \begin{equation}
        \nabla F(\mathbf{r}) |\mathbf{r}'| =
        (\nabla F(\mathbf{r}) \cdot \mathbf{r}')   \boldsymbol{\tau} + F(\mathbf{r})\boldsymbol{\tau}'
    \end{equation}
  Equivalently, it reads  
     \begin{equation} \label{A2}
      \nabla^\perp F(\mathbf{r})  \abs{\mathbf{r}'} = F(\mathbf{r})\boldsymbol{\tau}'
    \end{equation}
    where $ \nabla^\perp F(\mathbf{r}):=  \nabla F(\mathbf{r})   -
        (\nabla F(\mathbf{r}) \cdot \boldsymbol{\tau})  \boldsymbol{\tau} $ is the projected 
         gradient $\nabla F(\mathbf{r})$ onto the normal hyper-plane.  
        WLOG, assume $F>0$, then we have 
        $   \nabla^\perp   F(\mathbf{r})$   parallels to the direction  $   \boldsymbol{\tau}'$ and 
   $\|\nabla^\perp \log F(\mathbf{r}(t))\| =\kappa(t)$, where  the curvature $\kappa(t)=\frac{\abs{ \boldsymbol{\tau}'(t) }}{\abs{\mathbf
   {r}'(t)}}$.

    In our case of max-flux path,  $F(\mathbf{r})
    =\exp(\beta U(\mathbf{r}))$ 
    where $\beta>0$ is the inverse temperature. 
    Equation \eqref{A2} becomes
    $ \nabla^\perp  U(\mathbf{r}(t))  = \frac{1}{\beta}\kappa(t) \mathbf{n}(t)$ where
    $\mathbf{n}:=\boldsymbol{\tau}' / \abs{\boldsymbol{\tau}'}$ the unit vector along the curvature direction, which is orthogonal to the tangent $\boldsymbol{\tau}$.
    This special Euler-Lagrangian equation has been derived before \cite{Berkowitz1983}.
    At large $\beta$, the equation becomes the  first order necessary condition $\nabla^\perp  U(\mathbf{r}(t))=0$  for the MEP  \cite{Berkowitz1983,TempNEB2003}.

\bibliographystyle{siamplain}
\bibliography{gad,DL_ref}
\end{document}